\documentclass[conference]{IEEEtran}
\let\labelindent\relax
\usepackage{amsmath,amssymb,amsfonts}
\usepackage{textcomp}
\usepackage[linewidth=1pt]{mdframed}
\usepackage{booktabs} 
\usepackage{listings}
\usepackage{xcolor}
\usepackage[linesnumbered,ruled,vlined]{algorithm2e}
\usepackage{booktabs} 
\usepackage{listings}
\usepackage{xcolor}
\usepackage{graphicx}
\usepackage{url}
\usepackage{amsmath}
\usepackage[justification=centering]{caption}
\usepackage{multirow}
\usepackage{lscape}
\usepackage{xspace}
\usepackage[T1]{fontenc}
\usepackage[scaled=0.75]{beramono}
\usepackage{marvosym}
\usepackage{pifont}
\usepackage{float}
\usepackage{tikz}
\usepackage{comment}
\usepackage{enumerate}
\usepackage{hyperref}
\usepackage{framed}
\usepackage{color}
\usepackage{enumitem}
\usepackage[bottom]{footmisc}
\usepackage{subfigure}
\ifCLASSINFOpdf
\else
\fi
\begin{document}
%
\title{V-Gas: Generating High Gas Consumption Inputs to Avoid Out-of-Gas Vulnerability}



%
\author{\IEEEauthorblockN{Fuchen Ma\IEEEauthorrefmark{1},
Ying Fu\IEEEauthorrefmark{1},
Meng Ren\IEEEauthorrefmark{1}, 
Wanting Sun\IEEEauthorrefmark{2},
Houbing Song\IEEEauthorrefmark{5},
Yu Jiang\IEEEauthorrefmark{1},
Jun Sun\IEEEauthorrefmark{4} and
Jiaguang Sun\IEEEauthorrefmark{1}}
\IEEEauthorblockA{\IEEEauthorrefmark{1}Tsinghua University}

\IEEEauthorblockA{\IEEEauthorrefmark{3}Nanjing University of Aeronautics and Astronautics}
\IEEEauthorblockA{\IEEEauthorrefmark{4}Singapore Management University}
\IEEEauthorblockA{\IEEEauthorrefmark{5}Embry-Riddle Aeronautical University}
\IEEEauthorblockA{\IEEEauthorrefmark{2}Baidu Online Network Technology (Beijing) Co., Ltd}}



\maketitle

\begin{abstract}
The out-of-gas error occurs when smart contract programs are provided with inputs that cause excessive gas consumption, and would be easily exploited to make the DoS attack. Multiple approaches have been proposed to estimate the gas limit of a function in smart contracts to avoid such error. However, under estimation often happens when the contract is complicated. 
In this work, we propose V-Gas, which could automatically generate inputs that maximizes the gas cost and reduce the under estimation cases. V-Gas is designed based on feedback-directed mutational fuzz testing. First, V-Gas builds the gas weighted control flow graph (CFG) of functions in smart contracts. Then, V-Gas develops gas consumption guided selection and mutation strategies to generate the input that maximize the gas consumption. 
For evaluation, we implement V-Gas based on js-evm, a widely used ethereum virtual machine written in javascript, and conduct experiments on 736 real-world transactions recorded on Ethereum. 44.02\% of the transactions would have out-of-gas errors under the estimation results given by solc, means that the recorded real gas consumption for those recorded transactions is larger than the gas limit value estimated by solc. While V-Gas could reduce the under estimation ratio to 13.86\%. Compared with other well-known feedback-directed  fuzzing engines such as PerFuzz and SlowFuzz, V-Gas can generate a same or higher gas estimation value on 97.8\% of the recorded transactions with less time, usually within 5 minutes. 
Furthermore, V-Gas has exposed 25 previously unknown out-of-gas vulnerabilities in those widely-used smart contracts, 5 of which have been assigned unique CVE identifiers in the US National Vulnerability Database.
\end{abstract}


%
\section{Introduction}

In order to avoid excessive consumption of resources, users will be charged with gas to execute smart contracts \cite{dannen2017introducing}. Gas cost is determined by the amount of resources consumed by computation and storage. \cite{BlockGeek}\cite{whatisgas}. 
Meanwhile, users may set a gas limit for each transaction to prevent potential malicious gas consumption. But if a transaction needs more gas than the limit set by the user, an out-of-gas error occurs and the transaction would fail. As a result, the transaction would be reverted and the gas spent on executing the transaction is wasted. 
Malicious users may provide specially crafted inputs to trigger a high-gas cost transactions to launch a Denial-of-Service(DoS) attack \cite{TeXFAQ}. Indeed, the out-of-gas error happens frequently. We investigated transactions made from 2:00 to 3:00 on May 1st, 2019 on Ethereum and found that 21 of them have out-of-gas errors. Gas waste caused by these failed transactions accounts for 5.64\% of all gas consumption. 


Extensive research has focused on how to estimate the gas limit of smart contract functions to avoid this error. For example, Solc \cite{solc}, a widely-used compiler for smart contracts written in solidity language, provides a gas cost estimating tool to give a predication on the gas consumption for each function. It statically analyzes each contract function and provides an estimation based on the opcodes contained in the function. Marescotti et al. \cite{marescotti2018computing} present a symbolic execution method to estimate gas cost. The inputs triggering the gas cost could be achieved by solving the equations generated by the constraints for each program path. Although those works solve the problem to some extent, but in practice, their reference value is limited and users have to set the gas limit with manually evaluation. For instance, solc often estimates the gas consumption to be `infinite' or under estimated values. The path conditions of symbolic based estimators are often complex to be solved and most of the results are incorrect.


In fact, finding out how many gas units used in a transaction is hard before the transaction is really executed. There are mainly two challenges. 
First, the gas consumption of a transaction comes from not only the gas charged by the opcodes, but also the gas used in data storage. It is usually not easy to estimate the data storage with static analysis and solve the complex data structure with symbolic techniques. 
Furthermore, a function which contains many branches may  behave differently under different block states or with different inputs. It is often infeasible to generate some inputs manually which could consume some certain units of gas. 



In this paper, we propose V-Gas, a method that can automatically generate  inputs which could lead to a high gas consumption of contract functions and reduce the under estimation ratio of existing works. 
V-Gas is designed based on feedback-directed fuzz testing. Specifically, V-Gas uses three steps to generate inputs: 1) Weighted control flow graph (W-CFG) generation, 2) Feedback-directed selection and mutation, 3) Contract execution. 
The first step generates a W-CFG for the contract, where each node in the CFG has a weight representing the gas cost of the node. 
The second step selects the seeds based on the feedback information including the coverage and gas consumption, and defines several mutators for selected seeds. 
The third step executes the functions with the mutated seeds and calculates the gas consumption of nodes and edges in W-CFG. 
The key idea behind V-Gas is that the goal of finding out-of-gas vulnerabilities can be posed as an optimization problem whose goal is to find a set of inputs that maximizes gas consumption of a target smart contract function.




For evaluation, we implemented V-Gas based on js-evm, one of the widely-used Ethereum Virtual Machine (EVM) for the execution of smart contracts. We conduct experiments on 1000 real-world transactions recorded on Ethereum, where contracts used by 736 transactions could be compiled successfully. 
If we set the gas limit according to the limit value estimated by solc, 44.02\% of the transactions will have out-of-gas errors, meaning that the recorded real gas consumption value is larger than the estimated value. There are also 33.43\% of the transactions will be evaluated to be `infinite' by solc, which may mislead the users to set a higher gas limit and drag the execution process of the transaction. 
While V-Gas could reduce the under-estimation ratio to 13.86\%, and can also expose meaningful reference value for all the `infinite' excessive gas estimation situations by solc. We also compared V-Gas with other well-known feedback-directed fuzzing tools such as SlowFuzz \cite{slowfuzz} and PerFuzz \cite{Lemieux2018PerfFuzzAG}. The results show that V-Gas outperforms others. In particular, V-Gas can generate a same or higher gas estimation on 97.8\% of transactions with less time, usually within 5 minutes. Besides, V-Gas found 25 previous unknown gas-related vulnerabilities in widely-used real world contracts which could lead to a Denial-of-Service attack, 5 of which have been assigned unique CVE identifiers in the US National Vulnerability Database and other 20 are appended for approval. 
%
%
%
In general, this work makes the following contributions: 
\begin{enumerate}
\item We design and implement V-Gas, a feedback-directed fuzzing approach that could generate inputs triggering a high-gas cost of smart contracts, and reduce the under estimation ratio of existing works. 
\item We show two practical applications of V-Gas. First, V-Gas can be used to provide guidance on setting the gas limit to prevent potential out-of-gas errors. Second, V-Gas could be used to reveal gas-related vulnerabilities of existing contracts and settings. We have found 25 confirmed vulnerabilities, and 5 of which with CVE IDs.

\end{enumerate}

\indent The rest of the paper is organized as follows. In Section 2, we introduce the background of blockchain and provide a high-level overview of V-Gas with a motivating example. We formally describe the design of V-Gas in Section 3. Section 4 demonstrates the evaluation results. Section 5 represents some limitations of V-Gas. In Section 6, we introduce some other related works. Section 7 makes a conclusion.

\section{Overview}
\subsection{Background of Blockchain and Gas}

Blockchain is a distributed system widely used in all walks of life. This is a new application mode of computer technology including distributed data storage, point-to-point transmission, consensus mechanism, encryption algorithm and so on. Blockchain technology uses consensus mechanism to ensure that transactions are traceable. Meanwhile, the transactions on blockchain system could not be tempered with arbitrarily. In the aspect of privacy protection, blockchain technology uses many encryption algorithms to protect user's transaction data. 
%
Ethereum is a successful practice of blockchain system. In the original Bitcoin system, users could only make system-specified transactions, such as buying Bitcoin. In the Ethereum system, the user's freedom of action has been greatly improved. Developers can develop their own applications in Ethereum for publication. The applications deployed in the Ethereum is consists of smart contracts. Smart contract is written in Turing-completed language such as solidity. Ethereum also provides a virtual machine to execute the solidity code. The Ethereum Virtual Machine, also known as EVM, could recognize the binary code of the smart contracts as some opcodes, and execute the opcodes one by one.

\begin{figure}[!htbp]
\centering
\includegraphics[height=4.5cm, width=8.0cm]{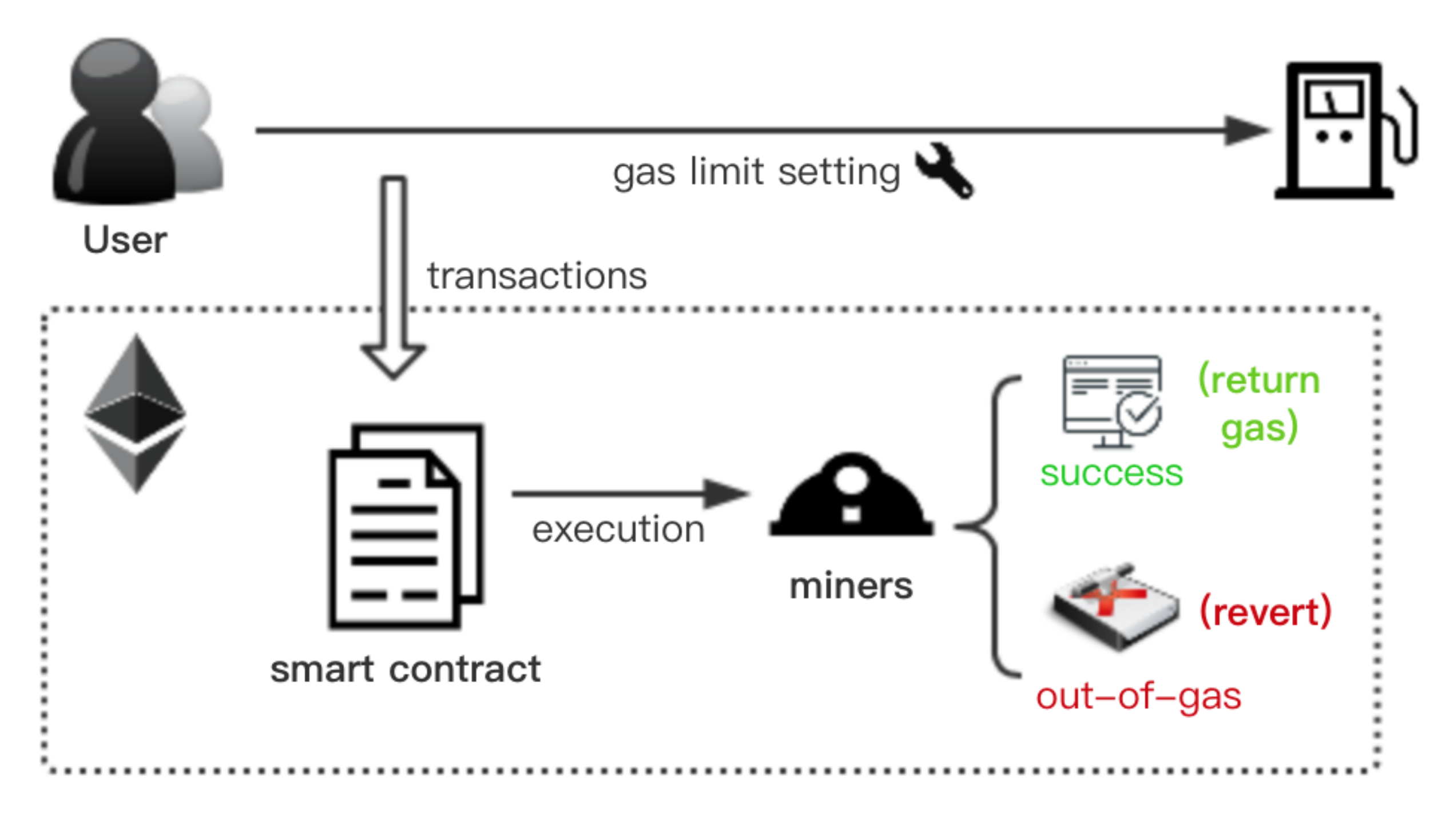}

\caption{Gas Mechanism in Ethereum, where miners will keep the gas though the transaction is not completed}
\label{gas_mech}
\end{figure}

Gas is a unit used to measure the workload of an operation in Ethereum. If an operation consumes more resources, it consumes more gas. This mechanism ensures that resources are not maliciously wasted. In the actual payment of gas costs, the amount of tokens consumed will be determined by the gas unit cost and the current price of gas. During the actual execution of the transaction, users will set up a maximum gas unit willing to pay for the transaction, which is called the gas limit. If the gas actually consumed by the transaction is larger than the gas limit, the execution of the entire transaction will fail, which is called an out-of-gas exception. Fig \ref{gas_mech} represents a work flow of gas mechanism in Ethereum.

It's not good if an out-of-gas error happens, because the miners will still charge the users for gas though the transaction was not completed. However, it is also bad to set a very high gas limit. Due to the limitation of block gas limit, the miners preferred transactions with a low gas limit. A high gas limit will drag on the speed of completing the transaction. Besides, a high gas limit setting, which allows more opcodes executing, may give some attackers a chance to commit an attack.

\subsection{A Motivating Example}
In this work, given a function in a smart contract, we aim to help users set gas limits more accurately, so as to avoid out-of-gas errors. There are two kinds of gas consumption mechanism in Ethereum, which are known as the execution gas cost and the transaction gas cost. The execution gas cost is based on the cost of computational operations which are executed as a result of the transaction. The transaction gas cost is based on the cost of sending data to the blockchain. There are 4 items which make up the full transaction cost: 1) the base cost of a transaction (which is 21,000 gas units); 2) the cost of a contract deployment (only charged on construct functions, 32,000 gas units); 3) the cost for every zero byte of data or code for a transaction; 4) the cost of every non-zero byte of data or code for a transaction. Given a trace of a smart contract, we can identify its gas cost by accumulating the gas consumption of each step. The problem is thus to identify the most gas-costing traces. Given that the gas cost is path-based property, as far as we know, existing methods (e.g., fuzzing with the objective of maximizing branch coverage, static analysis or symbolic execution) are not effective on solving the problem. So we try to design trace-based feedback guided fuzzing technique. 

In order to understand how our trace based technique works, let us consider a contract deployed on the Ethereum. The following code is intercepted from the contract \cite{motivating_example} which is used to distribute tokens. The function we focus on is \textit{distributeFixed}. This function transfers a fixed amount of ethers to each account that needs to be rewarded. Since this contract was deployed on Ethereum, 52 transactions have been made. Among these transactions, 80.77\% of them are executed successfully, and 15.38\% of the transactions failed because of out-of-gas error. All the out-of-gas transactions called the function \textit{distributeFixed}. The function is shown as below:

\definecolor{verylightgray}{rgb}{.97,.97,.97}

\lstdefinelanguage{Solidity}{
	keywords=[1]{anonymous, assembly, assert, balance, break, call, callcode, case, catch, class, constant, continue, constructor, contract, debugger, default, delegatecall, delete, do, else, emit, event, experimental, export, external, false, finally, for, function, gas, if, implements, import, in, indexed, instanceof, interface, internal, is, length, library, log0, log1, log2, log3, log4, memory, modifier, new, payable, pragma, private, protected, public, pure, push, require, return, returns, revert, selfdestruct, send, solidity, storage, struct, suicide, super, switch, then, this, throw, true, try, typeof, using, value, view, while, with, addmod, ecrecover, keccak256, mulmod, ripemd160, sha256, sha3}, 
	keywordstyle=[1]\color{magenta}\bfseries,
	keywords=[2]{address, bool, byte, bytes, bytes1, bytes2, bytes3, bytes4, bytes5, bytes6, bytes7, bytes8, bytes9, bytes10, bytes11, bytes12, bytes13, bytes14, bytes15, bytes16, bytes17, bytes18, bytes19, bytes20, bytes21, bytes22, bytes23, bytes24, bytes25, bytes26, bytes27, bytes28, bytes29, bytes30, bytes31, bytes32, enum, int, int8, int16, int24, int32, int40, int48, int56, int64, int72, int80, int88, int96, int104, int112, int120, int128, int136, int144, int152, int160, int168, int176, int184, int192, int200, int208, int216, int224, int232, int240, int248, int256, mapping, string, uint, uint8, uint16, uint24, uint32, uint40, uint48, uint56, uint64, uint72, uint80, uint88, uint96, uint104, uint112, uint120, uint128, uint136, uint144, uint152, uint160, uint168, uint176, uint184, uint192, uint200, uint208, uint216, uint224, uint232, uint240, uint248, uint256, var, void, ether, finney, szabo, wei, days, hours, minutes, seconds, weeks, years},	
	keywordstyle=[2]\color{teal}\bfseries,
	keywords=[3]{block, blockhash, coinbase, difficulty, gaslimit, number, timestamp, msg, data, gas, sender, sig, value, now, tx, gasprice, origin},	
	sensitive=false,
	morecomment=[l][{\color[RGB]{127,140,141}}]{//},
	morecomment=[s][{\color[RGB]{127,140,141}}]{/*}{*/},
	commentstyle=\color{codegreen}\ttfamily,
	stringstyle=\color{black}\ttfamily,
	morestring=[b]',
	morestring=[b]"
}
\begin{lstlisting}[
	label={list:motivating_example},
	language={Solidity},
    ]
contract DistributeTokens is Ownable{
  ...
  function distributeFixed
  (address[] _addrs, uint _amoutToEach) onlyOwner{
  /* _addrs is an array of address input by the user */
    for(uint i = 0; i < _addrs.length; ++i){
    /* transfer some token to each address */
     tokenReward.transfer(_addrs[i],_amoutToEach);
    }
  }
  ...
}
\end{lstlisting}

\begin{figure}[!htbp]
\centering
\includegraphics[height=4cm, width=8.0cm]{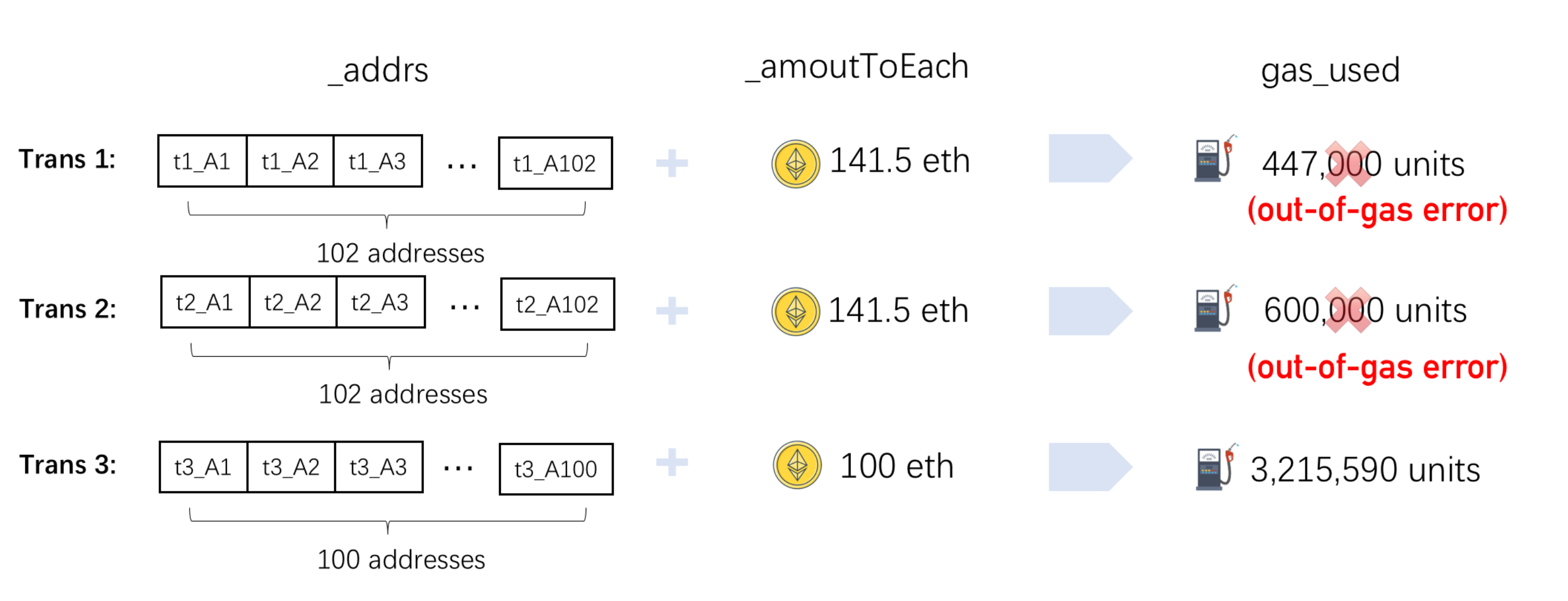}
\caption{Inputs for distributeFixed}
\label{input}
\end{figure}

If we use the gas estimation tool provided by solc to estimate the gas cost of this function, the result is `infinite'. The main source of gas consumption in \textit{distributeFixed} is the transfer at line 5. The transfer function is implemented based on the opcode CALL. CALL has a multi-part gas cost: 1) the basic gas cost is 700 units; 2) If the transferring value is non-zero, another 9000 gas units are charged; 3) If the destination account does not yet exist, another 25000 gas units are charged. This is the gas consumption of opcodes. In a real transaction, the memory used to store the transaction data on Ethereum also requires gas. 
We fetched 3 real transactions based on this function from Etherscan\cite{etherscan}. The inputs of the transactions and the results of gas used are shown in Fig. \ref{input}. 

The first transaction has 102 transfers. The gas limit set by users is 447,000 units. This transaction failed because of an out-of-gas error. The second transaction also has 100 ERC-20 transfers. The inputs of this transaction are the same as the first transaction. The user would like to try a higher gas limit for the transaction but significantly under-approximate the gas consumption and, as a result, the transaction also failed. The third transaction has 100 ERC-20 transfers from the contract account to 100 other accounts, the amount in the transaction is 100 gwei. The gas cost of the transaction with this set of input is 3,215,590 units. This transaction is successfully executed. The first two transactions cost 1,047,000 units of gas in total. According to the setting of gas price in these two transactions, 31,857,000 gwei is charged by these transactions. \textbf{That means nearly six dollars are wasted in vain after these transactions. These out-of-gas errors would not occur if the user knows in advance how many gas will be consumed by this function and what are the inputs to trigger the gas consumption. V-Gas addresses exactly this concern.}

\subsection{Gas Fuzzing Idea}

\begin{figure}[!htbp]
\centering
\includegraphics[height=7.5cm, width=8cm]{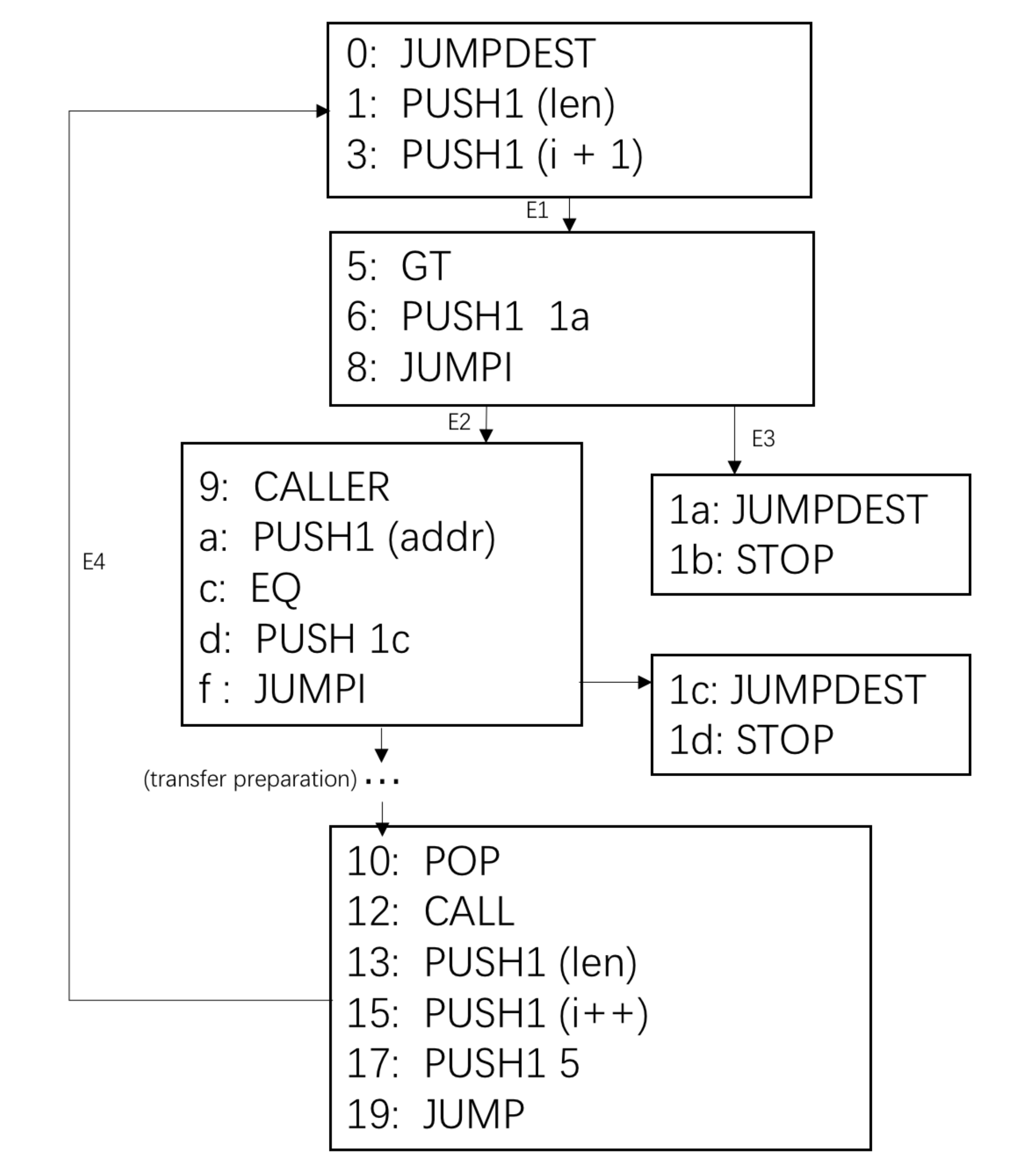}
\centering
\caption{The Control Flow Graph for function distibuteFixed}
\label{CFG}
\end{figure}

Let us look at the original control flow graph of the function \textit{distributeFixed} presented in Fig. \ref{CFG}. The transfer process needs many parameters, the complex sequences of opcodes are simplified in this figure to explain how V-Gas works. Because of the long trace as shown in the CFG, it is infeasible to use existing symbolic execution based tools to get an accurate value of gas consumption of this function. As mentioned above, static analysis based tools result in an estimation of `infinite'. In contrast, V-Gas applies a feedback-guided fuzzing techniques to identify an accurate gas estimation. The main idea of V-Gas is to reserve the seeds that could cost more gas units on some of the statements in a given function to trigger a high gas consumption.

We assume that the initial inputs of this function is the same as the third transaction in Fig. \ref{input}. V-Gas would set up an initial environment and execute the function with the initial inputs. The initial inputs would execute edge E1, E2 and E4 for 81 times. V-Gas will record the gas cost on each edge of the CFG as well as the total gas cost of the transaction. The initial inputs are reserved in the seeds pool for further mutation. At each iteration, V-Gas picks an input from the seeds pool and randomly mutate it with several mutators. 

\begin{figure*}[!htbp]
\includegraphics[height=8.0cm, width=18.0cm]{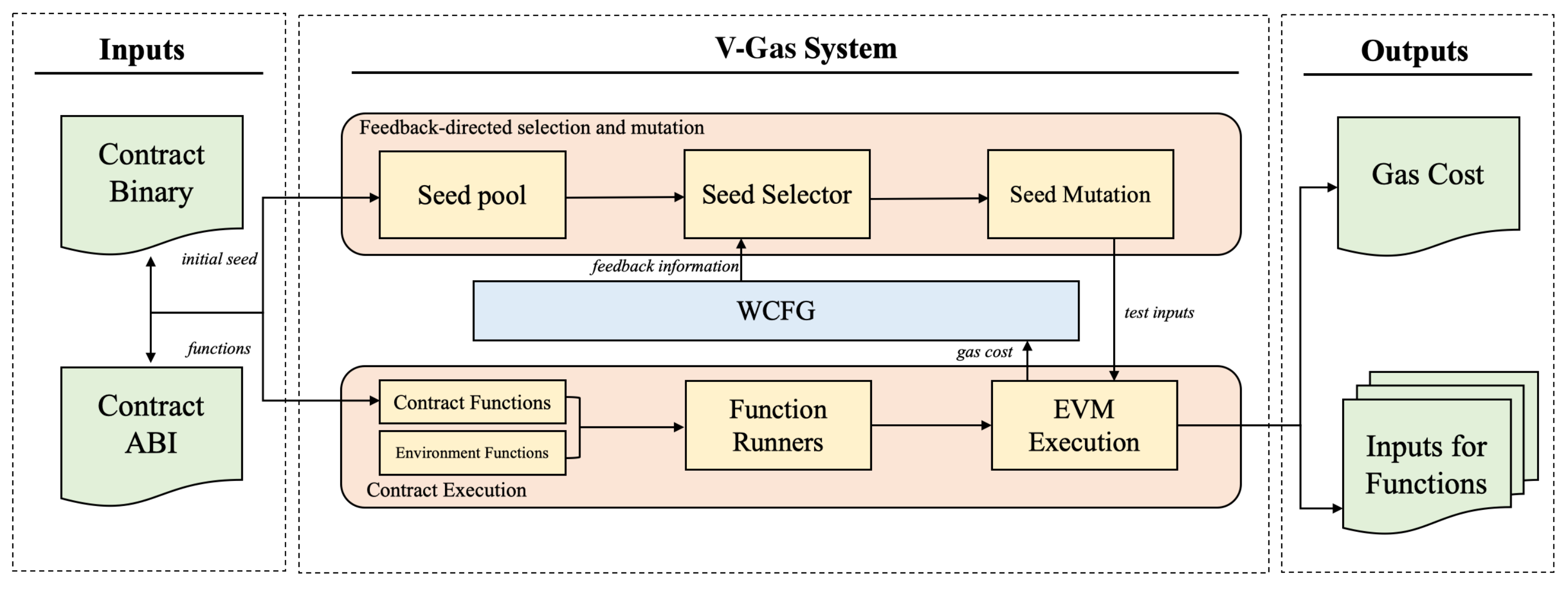}
\caption{V-Gas architecture, which consists of: W-CFG Generation, Mutation and Environment Setup. The inputs of V-Gas are the smart contract, while the outputs are the value of gas cost and the inputs that could trigger the cost.}
\label{architecture}
\end{figure*}

As Table \ref{V-Gas_motivating} shows, let us assume that the first mutated seed is \textit{Input2}. V-Gas will use the new input to run the function, the gas cost of this input only executes the edge E1, E2 and E4 for 32 times. The gas costs on these edges are not increased with the new input, and the total gas cost is not increased either, so it will be abandoned by V-Gas. Let us further assume that the next input is \textit{Input3}. With this input, the edges E1, E2 and E4 are executed for 100 times. The gas cost of these edges are increased, and this input will be reserved in the seeds pool by V-Gas. The gas cost for each edge and the total gas cost are updated. As for \textit{Input 4}, the edges are executed for 100 times as well, but if the total gas cost is increased, V-Gas will keep it into the seeds pool as well. All of the reserved inputs will be mutated to generate new inputs. From all of the alternative inputs, V-Gas will choose several ones which have the most gas cost of the whole function as the basic input to be mutated. In this way, V-Gas always keeps the inputs that could trigger a local higher gas cost or a global higher gas cost and makes mutations based on these inputs. 

\begin{table}[!htbp]
\caption{Overview of the input mutation for V-Gas}
\label{V-Gas_motivating}
\begin{center}
\begin{tabular}{c|ccc}
\hline
\multicolumn{1}{l|}{Input\_Num} & \multicolumn{1}{l}{\_addr.length} & \multicolumn{1}{l}{\_amount}  & \multicolumn{1}{l}{edge\_exe\_times} \\ \hline
Input1                           & 81                                 & 100 eth                        & 81                                    \\
{\color[HTML]{FE0000} Input2}    & {\color[HTML]{FE0000} 32}          & {\color[HTML]{FE0000} 231 eth} & {\color[HTML]{FE0000} 32}             \\ 
Input3                           & 100                                & 109 eth                        & 32                                    \\
Input4                           & 100                                & 31 eth                         & 32                                    \\ \hline
\end{tabular}
\end{center}
\end{table}

In this contract, if the length of the address array is larger, the transaction will make more transfer operations and consume more gas, eventually exceed the gas limit of a block which is 80,039,143 units \cite{Ethstat}. V-Gas generated sets of inputs that trigger a high gas cost which exceeds the gas limit. 
Malicious attackers may exploit this to implement an attack. The vulnerability has been assigned with a CVE Id CVE-XX-XX (hidden because of the double-blind review mechanism). A more thorough analysis of V-Gas's performance on various smart contracts will be presented in Section IV.

\section{V-Gas Design}

 V-Gas uses three steps to generate inputs to maxmize the gas cost: 1) Weighted control flow graph (W-CFG) generation, 2) Feedback-directed mutation and selection, 3) Contract execution. The architecture of V-Gas is shown in Fig\ref{architecture}. V-Gas starts the fuzzing process with a binary file (.bin) and application binary interface file (.abi) of the contract. V-Gas extracts the types of parameters and randomly generates some initial inputs for contracts and generates some initial environment variables. Besides, V-Gas  deploys the contract and prepares some function runners for execution. V-Gas uses the first input to run, and calculates the gas cost information for each edge in W-CFG. After the execution, the seed and feedback information are passed to the feedback-directed selection and mutation part. The feedback information determines whether the seed is saved or not. All the saved seeds are mutated under V-Gas mutation strategies. After the mutation, V-Gas selects one seed from the queue and pass it to the function runners. This is a cyclic process and V-Gas only stops when the user forces it to stop with time or value thresholds.

\subsection{W-CFG Generation}
\begin{figure}[!htbp]
\centering
\includegraphics[height=7.0cm, width=9.0cm]{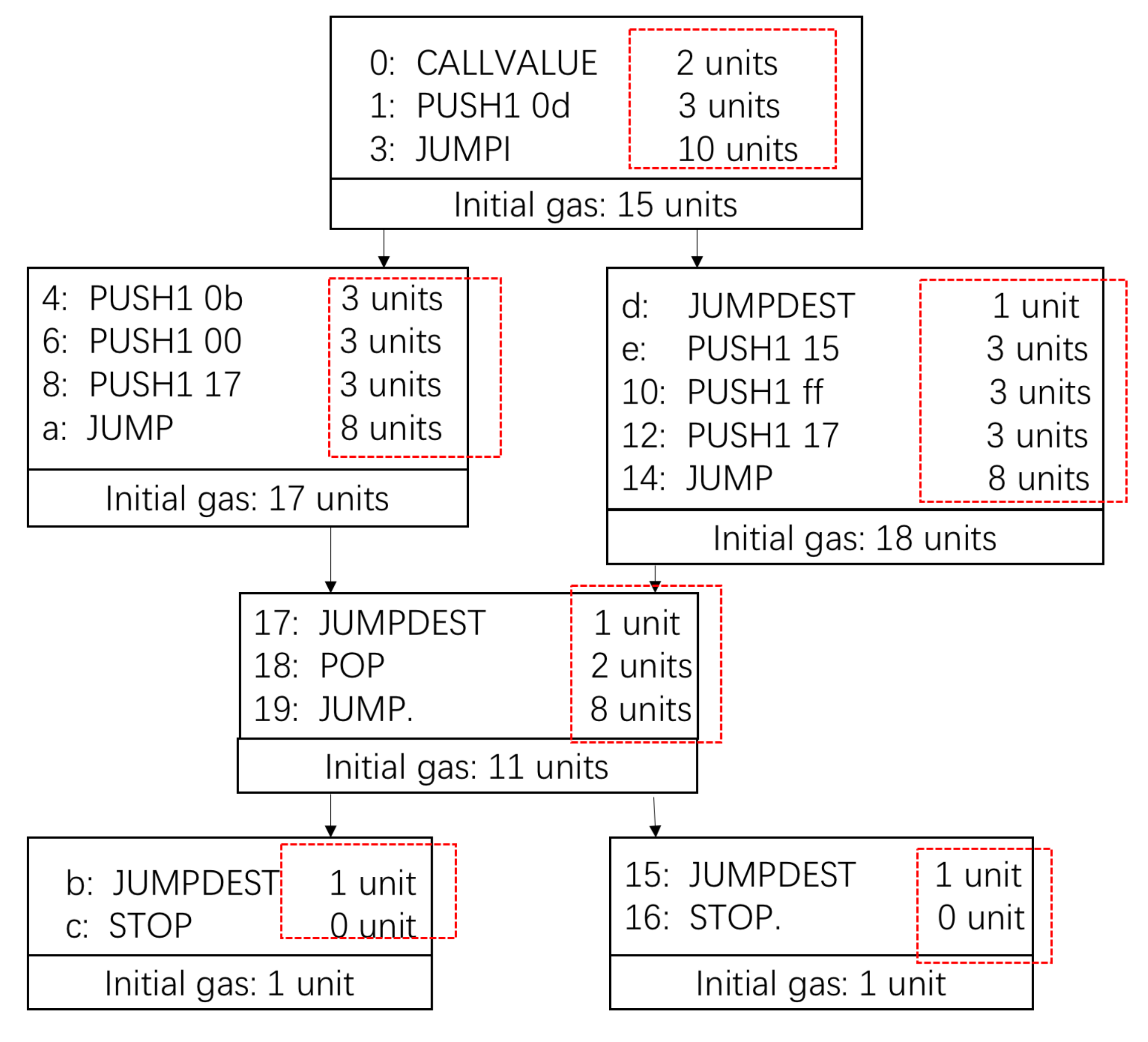}
\caption{Example of W-CFG, each block has an initial gas costs. When there is a JUMP or JUMPI opcode, there is a new branch.}
\label{WCFG}
\end{figure}

V-Gas defines a weighted control flow graph which gives a gas consumption weight to each node contained in the traditional CFG. Each node of the CFG is a sequential non-branching block of opcodes. If there is a JUMP opcode or JUMPI opcode, CFG will have a new branch. The weight of the node represents the gas cost of the node. As for the gas cost of some opcodes determined by the operands, a node may have an uncertain gas cost. If the node has an uncertain gas cost, weight of this node is defined as gas consumption values excluding uncertainties. A simple W-CFG generated by V-Gas is presented in Fig. \ref{WCFG}.

As we could see from the W-CFG presented above, each node has some opcodes, and an initial gas consumption value. If the gas consumption of a node is not fixed, the initial gas consumption of the node will be the minimum value of the node. The W-CFG has a new branch when encounter with a JUMP or JUMPI opcode. JUMP opcode takes one parameter from the stack as the destination. JUMPI opcode takes two parameters. The first parameter is the destination and the second one is the condition. As for this example, the destination of the JUMPI in the first node is `0d' which is pushed into the stack in the first PUSH opcode. The condition of this JUMPI is the msg.value which is pushed by CALLVALUE opcode. If msg.value is 0, JUMPI will add the program counter by 1. If msg.value is not 0, JUMPI will jump to address 17. Each edge in the W-CFG has an initial gas consumption of 0. The gas consumption of an edge is defined as the total units of gas cost of the root node of this edge in the execution process. 


\subsection{Feedback-directed selection and mutation}
Feedback-directed fuzzing is the kernel component of V-Gas. It keeps collecting feedback information during test case execution and using the feedback to optimize the fuzzing process, by selecting the right seeds for generating new test cases. Seeds selection determines whether a seed is a good one, and seeds mutation determines the how the selected seeds change for the next execution process. Algorithm \ref{feedback} outlines the feed-back directed selection and mutation for V-Gas. 
V-Gas will keep running until the duration time set by the user is up or the user would like to stop the process.  

\IncMargin{1em}
\begin{algorithm}
\SetAlgoLined
\SetKwData{Left}{left}
\SetKwFunction{Mutate}{Mutate}
\SetKwFunction{Run}{Run}
\SetKwFunction{push}{push}
\SetKw{Continue}{continue}
\SetKwInOut{Input}{Input}\SetKwInOut{Output}{Output}
\Input{initial\_inputs}
select\_seed = initial\_inputs\;
old\_feedBack = function.execute(select\_inputs)\;
SeedsPool.push(initial\_inputs)\;
\Repeat{Duration\_timeIsUp() || UserStop()}
{   
    \ForEach{\textrm{strategy} of \textrm{mutation\_strategies}}{
        newSeed = mutate(original\_seed, strategy)\;
        newSeed.change\_Array()\;
        new\_seeds.push(newSeed)\;
    }
    \ForEach{\textrm{seed} of \textrm{new\_seeds}}{
        feedBack = function.execute(seed)\;
        \uIf{feedBack.isInteresting(old\_feedfack)}{
            SeedsPool.push(seed)\;
            old\_feedback = feedBack\;
        }
    }
    select\_seed = select(SeedsPool)\;
}
\Output{Inputs that lead to a high gas cost or Warnings}
\caption{Feedback-directed selection and mutation process}\label{feedback}
\end{algorithm}\DecMargin{1em}


\textbf{Seeds Selection.} V-Gas could generate an initial input according to the types of parameters in smart contracts' functions. The initial inputs are put into the contract functions for the first execution. The feedback information of the execution, which contains gas cost and the number of times each edge is exercised, and the total gas cost, is stored by V-Gas. The initial seed is mutated then under several mutation operators. The mutated inputs are put into the function again. If the inputs could cost more gas of the function or hits more times of some edges in W-CFG, the inputs are defined as interesting seeds and would be reserved. The selection process could be represented as the following:

\[
\begin{split}
\{s|S, total(s)>totalcur\}         \cup \\
\{s|S,\exists e:E ~cost(s,e)>costcur(e)\}
\end{split}
\]
%
\\where $E$ represents the set of edges in the W-CFG, $s$ represents seeds, and \textit{total} represents the total gas cost of the transaction. 
All interesting seeds are stored in a priority queue called seeds pool. Seeds that consume more gas are assigned with higher priority. 
Each iteration, V-Gas picks a seed from the queue to mutate, and puts the mutated seed into the function again. 
V-Gas repeats the above process iteratively. 


For the local optimum problem, it occurs when a node does not has an growth on its gas consumption, which hinders the execution of its subsequent nodes which may lead to a high gas consumption. In order to tackle with this problem, V-Gas uses an sampling algorithm similar to MCMC \cite{MCMC} in seed selector which is implemented in the function \textit{isInteresting} in the algorithm above. In this way, V-Gas saves those none-interesting seeds with a probability and successfully avoid the local optimal problems.

\textbf{Seeds Mutation.} Seeds mutation operators of V-Gas could be divided in two groups: 1) {Traditional mutation.} For mutating common inputs of functions, we defined several mutators which are adopted from AFL \cite{AFL}. Each selected seed is mutated using the mutation operators. The mutators we adopt include bit flipping, byte flipping, arithmetic increment and decrement of integer values, replacing of bytes with interesting integer values (0, MAX\_INT), etc. 
2) {Gas-cost-specific mutation.} The second group of the mutation operators are used for mutating some special type of variables such as arrays. The length of an array variable may affect the times of the execution of some loops in a function. More loops means more gas consumption somehow. An array variable is replaced by a new array with a different length and different elements. The elements of the array are generated randomly according to the type of the parameters, and the length is randomly selected. 

\indent The seed in V-Gas is represented as a gene chain, which contains all bytes of the function inputs in one contract. V-Gas defines a gene\_map to match the inputs for each parameter of each function. Besides, smart contracts need some state information to run. V-Gas defines state variables as new functions and connects them with the original functions in the contract. The inputs of the functions are state variables. V-Gas will generate new state variables with the same strategies as other inputs and store them in the gene chain.

\subsection{Contract Execution}
The input of V-Gas is the abi file and binary file of a smart contract. Construction of an environment and several runners for a contract to run is essential. The environment of V-Gas is a simulation one which contains the same features as the Ethereum environment. For each contract, the first step is to deploy the contract, that is, to execute the construct function of the contract. The construct process is done by executing the binary code compiled from the contract source code with on the Ethereum virtual machine. After one contract has been deployed, V-Gas generates a runner for each function. This runner contains the signature of the function and each input for the function. With the help of the runner, V-Gas will find the matching piece of bytecode from the bin file according to the signature of the function.  The signature of a function is a string that only related to the name and the inputs of a function. The number of inputs and their types are extracted by V-Gas from the abi file, and recorded in the runner of function. Ethereum virtual machine could run a transaction easily with the parameters provided by these drivers. To startup the fuzzing process, an initial seed is also needed. The initial seed is generated by the abi file of the contract. A map is used to record the mapping relation between the formal parameter and the actual parameter stored in the seed.

For detail implementation, V-Gas is based on js-evm, and relays on several open-source programs such as solidity-types\cite{SolidityTypes}, to accomplish the whole task. 
types to generate a variable for a certain type in smart contracts. 
In the following, we mainly describe some implementation details for mutation part and environment variables definition part.

\indent The seed is represented as a variable called gene in our implementation. A gene stores all of the inputs of the functions in one contract. Another structure named gene\_map is used to fetch seeds for each input parameter of a function. For each input of the function defined in the contract, V-Gas fetches the name of the function, the length of the inputs, the name of the input and the type of the input as the components to generate a key. The key is used to get the seed from the gene\_map. For each input, we use an open-source tool called 'solidity-types' to generate a random value according to the type of the parameter. The value is represented as a buffer and connected to the variable gene. The starting position and the ending position of the value in the gene are recorded in the gene\_map. By this way, we generate random values for the inputs and maintain a structure to effectively get the values.
%
As for the environment variables, we define the following variables in V-Gas: the coinbase of a block as an address variable; the difficulty of a block as an integer; the block number as an integer; the block timestamp as an integer; the sender as an address; the original address of a transaction named tx.origin as an address. 

\section{Evaluation}
In this section, we evaluate V-Gas to answer the following four questions: \textbf{a) Is out-of-gas a real threat? b) Is V-Gas capable of generating inputs that approximate the gas limit for a given smart contract function?} \textbf{c) What advantages does V-Gas have compared with other feedback-directed fuzzing tools?}
\textbf{d) Is V-Gas able to be applied in the practical use?}

\indent All our experiments were performed on a machine with 126GB memory, equipped with an Intel(R) Xeon(R) Gold 6148 CPU @ 2.40GHz and running 64-bit Linux operating system. All of the 255 real world contracts used in our experiment are compiled by solc version 0.4.26. We took 1000 recorded transactions randomly from Etherscan\cite{etherscan} for all 
of the following experiments. We used solc to compile all these contracts and contracts used by 736 transactions could be compiled successfully. The other contracts could not be compiled because of the version of the compiler. It's necessary to mention here that some of the tools referred in related work are not available at present, so we will not compare V-Gas with them in the following experiments, but with the most widely used solc and reward-directed fuzzing such as SlowFuzz and PerFuzz.

\subsection{Risks of out-of-gas error}
\begin{figure}[h]
\centering
\includegraphics[height=6cm, width=9.0cm]{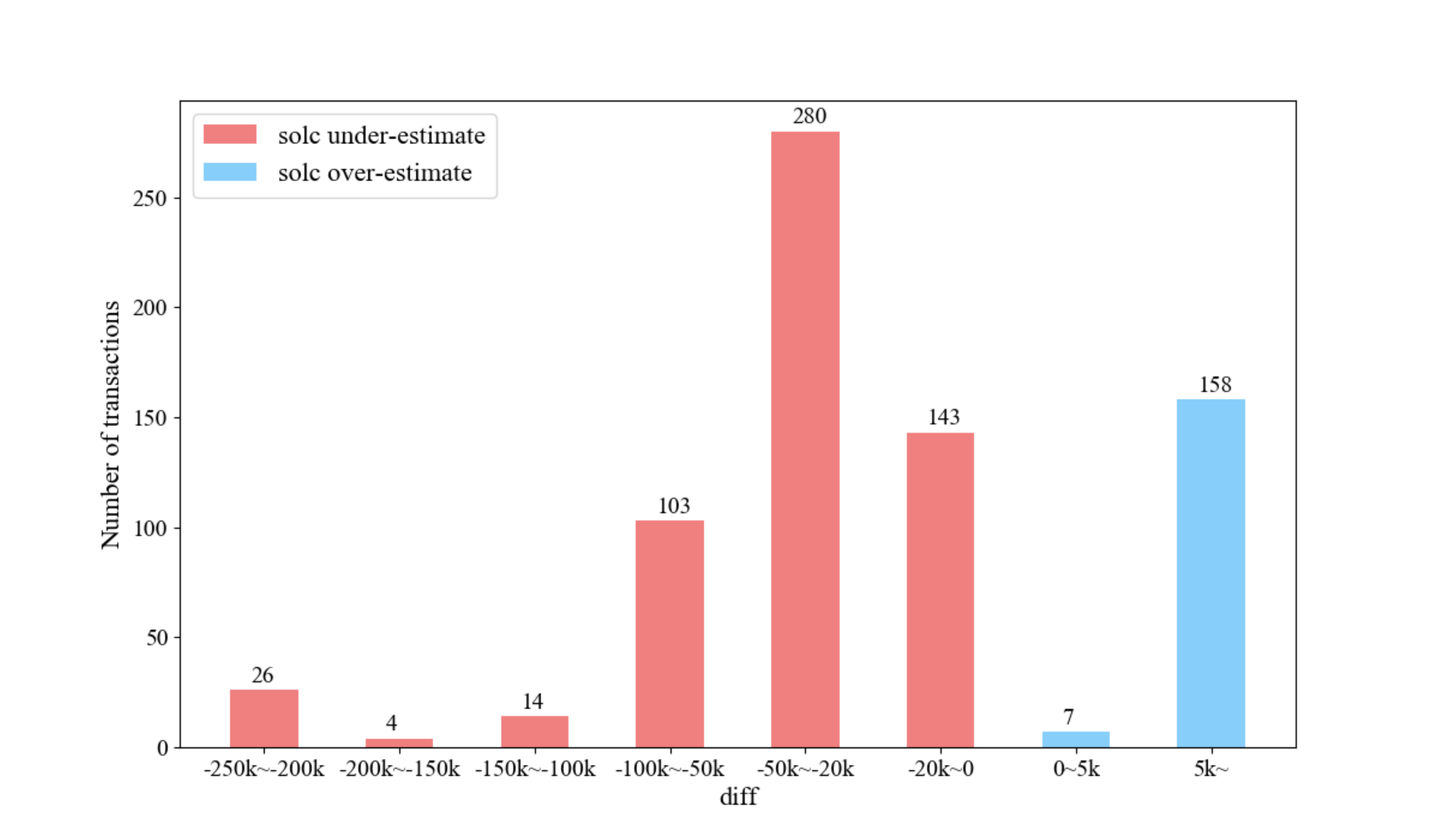}
\caption{Risks of out-of-gas error with the results of solc}.
 \label{risks}
\end{figure}

Because of the fact that lots of users of smart contracts use the estimation tool of solc to set the gas limit when they made a transaction, we use the result of solc to investigate the current level of out-of-gas risk. Fig.\ref{risks} shows a comparison between the gas used of these transactions and the gas estimation results of the tool in solc. 
We use `diff' to express this contrast. As shown in the following formula: 
\[
\begin{split}
\ diff = solc_{est}(function) - real\_cost(trans) \
\end{split}
\]
where `diff' represents the difference between the gas estimation of solc and the gas units actually consumed by the recorded transaction. In Fig \ref{risks}, the X-axis represents the interval where `diff' is located, and the Y-axis represents the number of transactions in that interval. The red bar represents the case where the `diff' is negative, and the blue bar represents the case where the 'diff' is positive. We use 0 to represent the `infinite' estimation given by solc. \textbf{As the result shows, if the users use the estimation results of solc to set the gas limit, out-of-gas error is likely to happen. In particular, 44.02\% of the transactions will be out-of-gas, and 33.43\% of the transactions will evaluated to be `infinite' without meaningful guidance. }

{We chose the most representative 8 functions from the 255 contracts involved in the transactions as illustrative examples to show the detail situations.} 
The 8 functions are with the top most probability of out-of-gas error and are from different contracts. The functions' details and the gas estimation results of them are listed in Table \ref{contracts}. The column \textit{Trans} represents the number of transactions that called this function. \textit{solc} column shows the estimation result of solc. Column \textit{real} represents the recorded maximum cost of the executed transactions based on this function.
\begin{table}[H]
\caption{Chosen functions' details, and their estimation of gas cost by solc and recorded real gas cost.}
\label{contracts}
\begin{center}
\begin{tabular}{c|cccc}
\hline
Contract        & Function                                                          & Trans & solc     & real \\ \hline
BAToken       & transfer
    & 1       & 43517    & 21916    \\ 
DSTokenBase     & transfer                                                          & 1       & 43919    & 37175    \\ 
Dogg            & transfer                                                          & 2       & 45216    & 52835    \\ 
AccessoryData   & addSERAPHIM	  
    & 1       & 41921    & 49816    \\ 
ENIGMA          & Activate                                                          & 5       & infinite & 25689    \\ 
KittyItemMarket & additem                                                           & 8       & infinite & 113844   \\ 
CommunityChest  & send                                                              & 8       & infinite & 32523    \\ 
Future1Exchange & \begin{tabular}[c]{@{}c@{}}admin\_token\_\\ withdraw\end{tabular} & 4       & infinite & 31484    \\ \hline
\end{tabular}
\end{center} 
\end{table}
If users set gas limit according to the results of solc, 93.3\% transactions based on the 8 contracts will have out-of-gas problems or infinite limit. 
The under estimation is mainly because the gas estimation tool in solc uses static analysis to calculate the gas consumption of all opcodes. solc does not consider the gas consumption caused by data storage during the execution of transactions, which results in the estimation bias compared with the real situation. Furthermore, if the function has some complex loops which depends on the inputs for the function, the execution times of each opcode are uncertain. In this case, solc will give an estimate of 'infinite', and users have to set the gas limit with manually analysis, which is not correct. As mentioned in the later experiments, we detect out-of-gas vulnerabilities in 25 contracts out of the 255 tested contracts. 

\begin{mdframed}
\textbf{Answer to RQ-a:} The results of current gas estimation tools can lead to a large number of out-of-gas occurrences. Out-of-gas error is indeed a potential threat. 
\end{mdframed}

\subsection{V-Gas Performance}

\begin{figure}[hbp]
\centering
\includegraphics[height=6.0cm, width=9.0cm]{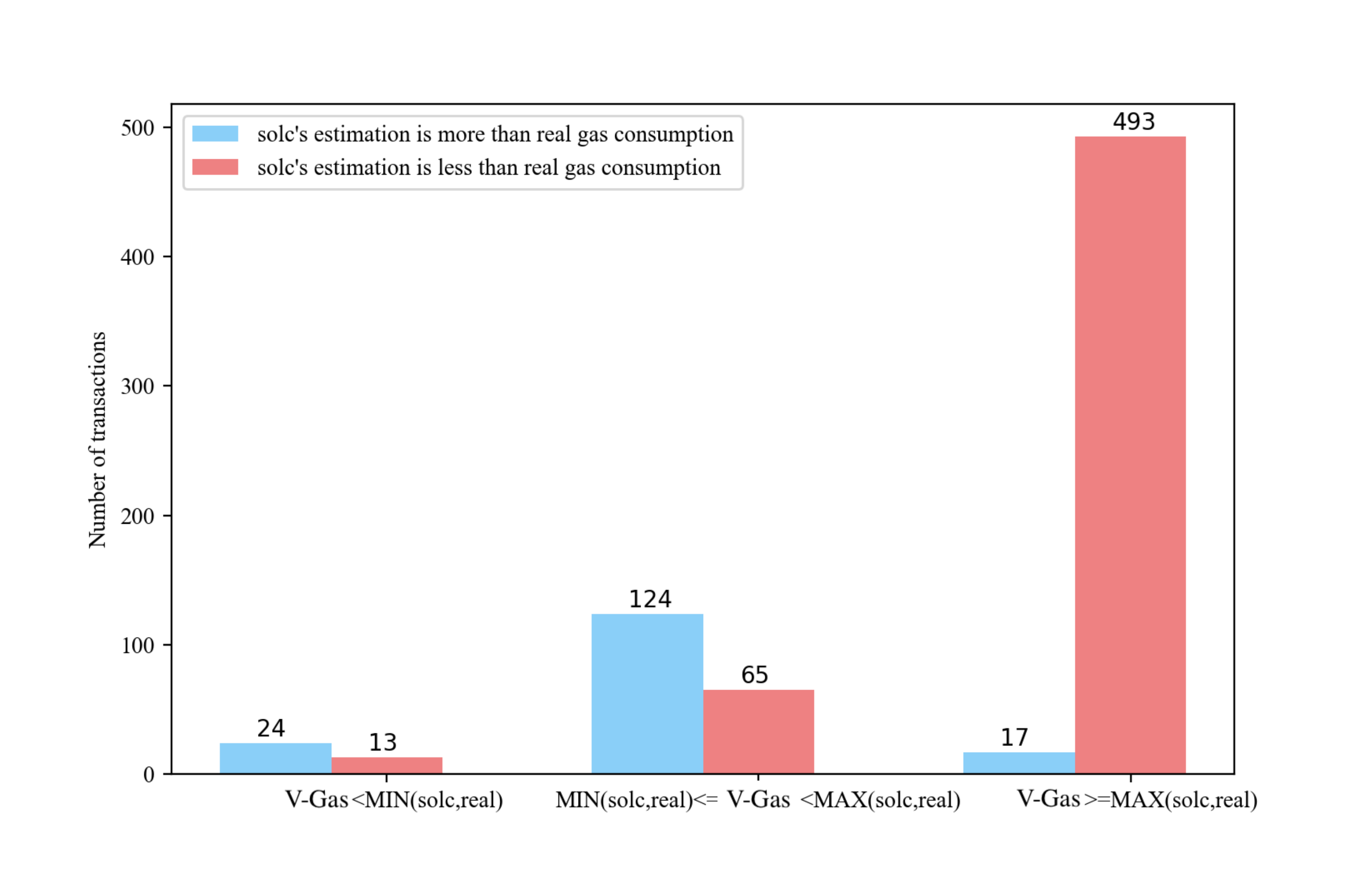}
\caption{V-Gas results compared with results of solc and the real gas consumption in the transaction}
\label{V-GasRealWorld}
\end{figure}

In order to answer the second question, we evaluated V-Gas with those functions  and ran each function with 5 minutes. The results are shown in Figure \ref{V-GasRealWorld}. According to the relationship among the results of V-Gas and solc and real gas cost, we divide the transactions into three intervals. As shown in Fig. \ref{V-GasRealWorld}, the first interval represents the results of V-Gas are smaller than solc and real gas cost, the second interval represents the results of V-Gas are between solc's result and real gas cost, and the third one represents the results of V-Gas are larger than solc and real gas cost. The red bars in the figure represent that, within this interval, solc's result is smaller than the real gas consumption and would definitely result in out-of-gas error, the blue bar represents that although the solc's result is larger than real consumption, but would still be potentially attacked by the inputs generated by V-Gas, especially in the third interval.

The figure shows that V-Gas can exceed the actual gas cost in 634 transactions (493+17+124) of the 225 contracts, counts for 86.14\% of all transactions. 
In these transactions, if user sets gas limit according to results of V-Gas not the solc, out-of-gas problems might be successfully avoided. As we illustrated in the formal section, the results of solc exceed the real gas consumption on only 22.55\% of the transactions. \textbf{If V-Gas is used for gas limit settings, the risk of out-of-gas occurring in these transactions will be greatly reduced, from 44.02\% to 13.86\%.}

We use the 8 illustrative examples in the formal subsection to observe the effectiveness of V-Gas. The results are shown in Table \ref{Ability}. Column V-Gas represents the results of maximum gas given by V-Gas in 5 minutes. Column Time shows how long it takes for V-Gas to find the maximum gas cost. The last two columns show the increase ratio of gas cost given by V-Gas compared with solc and the real situation.

\begin{table}[h]
\caption{V-Gas results on the 8 functions}
\label{Ability}
\begin{center}
\begin{tabular}{c|cccc}
\hline
Contract                               & V-Gas                       & Time                         & /solc                          & /real                          \\ \hline
{\color[HTML]{000000} BAToken}         & {\color[HTML]{000000} 25879}  & {\color[HTML]{000000} 0.542} & {\color[HTML]{000000} -40.5\%} & {\color[HTML]{000000} 18.08\%}  \\
{\color[HTML]{000000} DSTokenBase}     & {\color[HTML]{000000} 53179}  & {\color[HTML]{000000} 1.949} & {\color[HTML]{000000} 21.08\%} & {\color[HTML]{000000} 43.05\%} \\
{\color[HTML]{000000} Dogg}            & {\color[HTML]{000000} 55072}  & {\color[HTML]{000000} 0.583}  & {\color[HTML]{000000} 21.80\%} & {\color[HTML]{000000} 4.23\%}  \\
{\color[HTML]{000000} AccessoryData}   & {\color[HTML]{000000} 49601}  & {\color[HTML]{000000} 0.224} & {\color[HTML]{000000} 18.32\%} & {\color[HTML]{000000} -0.43\%} \\
{\color[HTML]{000000} ENIGMA}          & {\color[HTML]{000000} 26949}  & {\color[HTML]{000000} 8.496}  & {\color[HTML]{000000} /}       & {\color[HTML]{000000} 4.90\%}  \\
{\color[HTML]{000000} KittyItemMarket} & {\color[HTML]{000000} 116324} & {\color[HTML]{000000} 0.417} & {\color[HTML]{000000} /}       & {\color[HTML]{000000} 2.18\%}  \\
{\color[HTML]{000000} CommunityChest}  & {\color[HTML]{000000} 32884}  & {\color[HTML]{000000} 0.173} & {\color[HTML]{000000} /}       & {\color[HTML]{000000} 1.11\%}  \\
{\color[HTML]{000000} Future1Exchange} & {\color[HTML]{000000} 32991}  & {\color[HTML]{000000} 0.389}  & {\color[HTML]{000000} /}       & {\color[HTML]{000000} 4.79\%}  \\ \hline
\end{tabular}
\end{center}
\end{table}

We could find in the table that, V-Gas could generate inputs to trigger a gas which is higher than the estimation result of solc by 20\% on average. Except for the contract AccessorData, V-Gas's results are higher than the maximum value of real transactions. For the contract BAToken, V-Gas gives a lower result than solc, we will explain the reason below. For the contract AccessoryData, the gas consumption is lower than the real value by only 0.43\% which means V-Gas's result is very closed to the real value. If the user sets the gas limit with the estimation result by solc, 7 of the contracts are in dangerous of out-of-gas error. However, with V-Gas's result, only one of the contract may occur out-of-gas error. 

\begin{figure*}[htbp]
\begin{minipage}{0.24\linewidth}
  \centerline{\includegraphics[width=4.0cm]{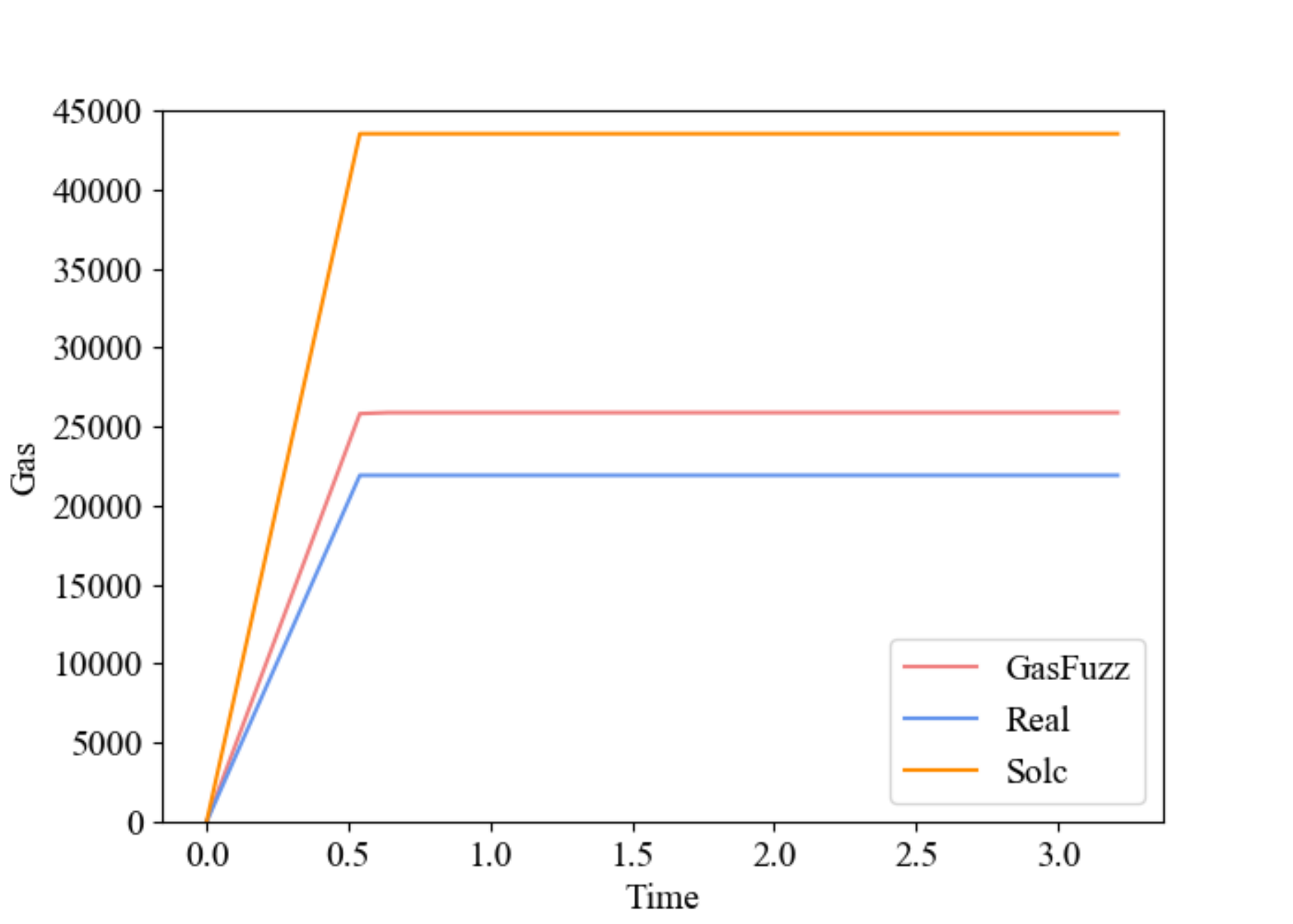}}
  \centering
  \small{(a) V-Gas on \\ BAToken.sol}
\end{minipage}
\hfill
\begin{minipage}{0.24\linewidth}
  \centerline{\includegraphics[width=4.0cm]{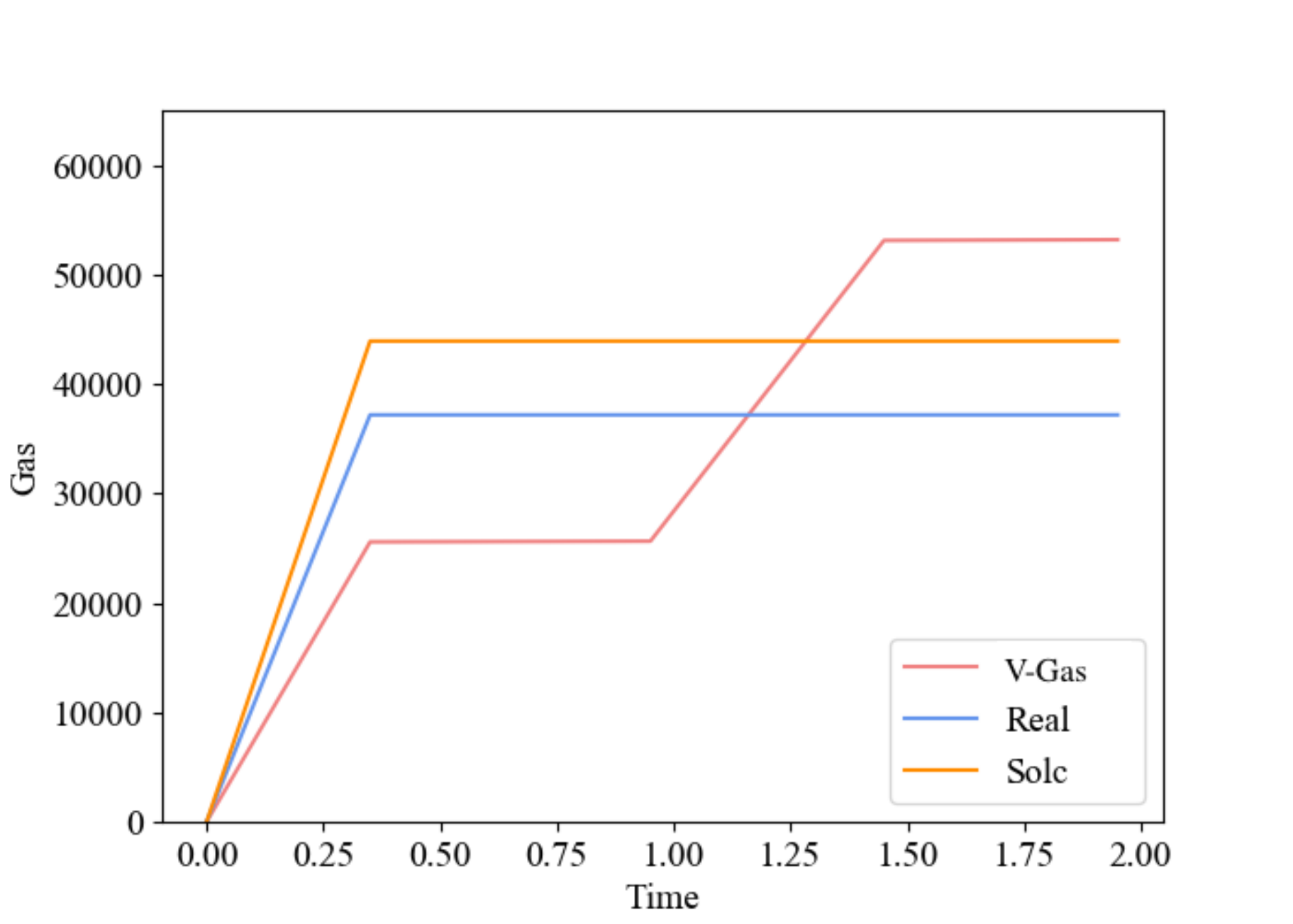}}
  \centering
  \small{(b) V-Gas on \\ DSTokenBase.sol}
\end{minipage}
\hfill
\begin{minipage}{0.24\linewidth}
  \centerline{\includegraphics[width=4.0cm]{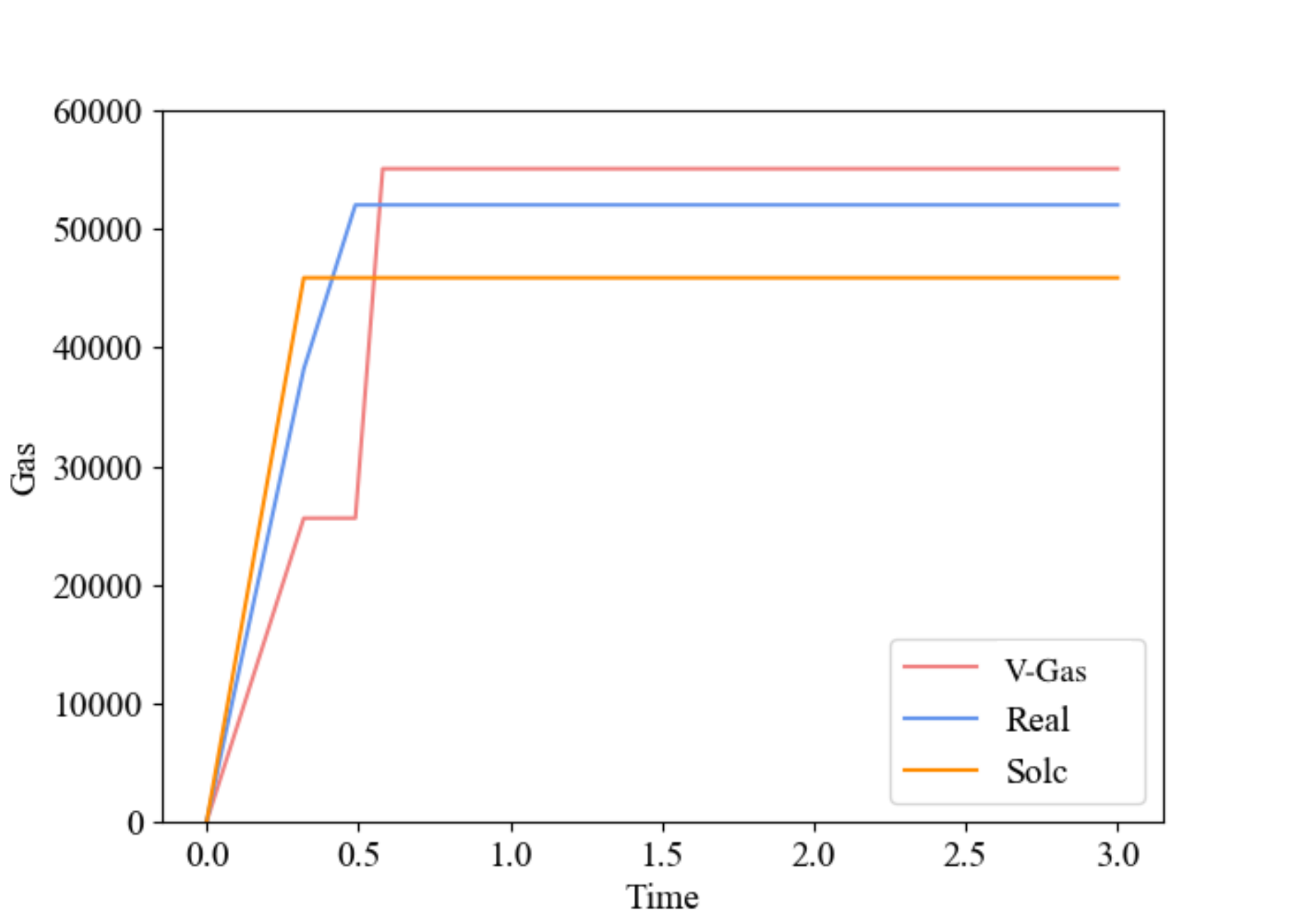}}
  \centering
  \small{(c) V-Gas on \\ Dogg.sol}
\end{minipage}
\hfill
\begin{minipage}{0.24\linewidth}
  \centerline{\includegraphics[width=4.0cm]{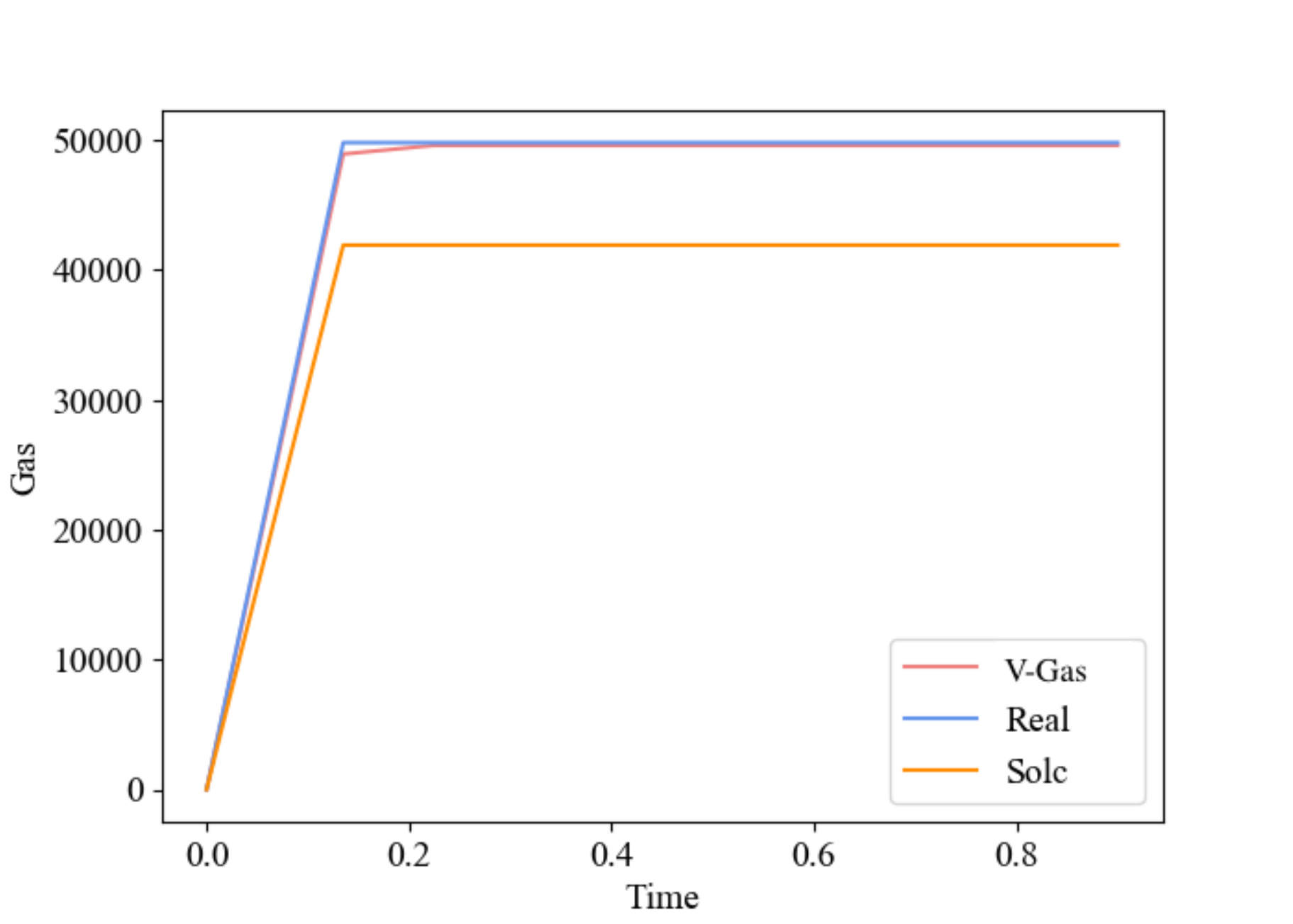}}
  \centering
  \small{(d) V-Gas on AccessoryData.sol}

\end{minipage}
\vfill
\begin{minipage}{0.24\linewidth}
  \centerline{\includegraphics[width=4.0cm]{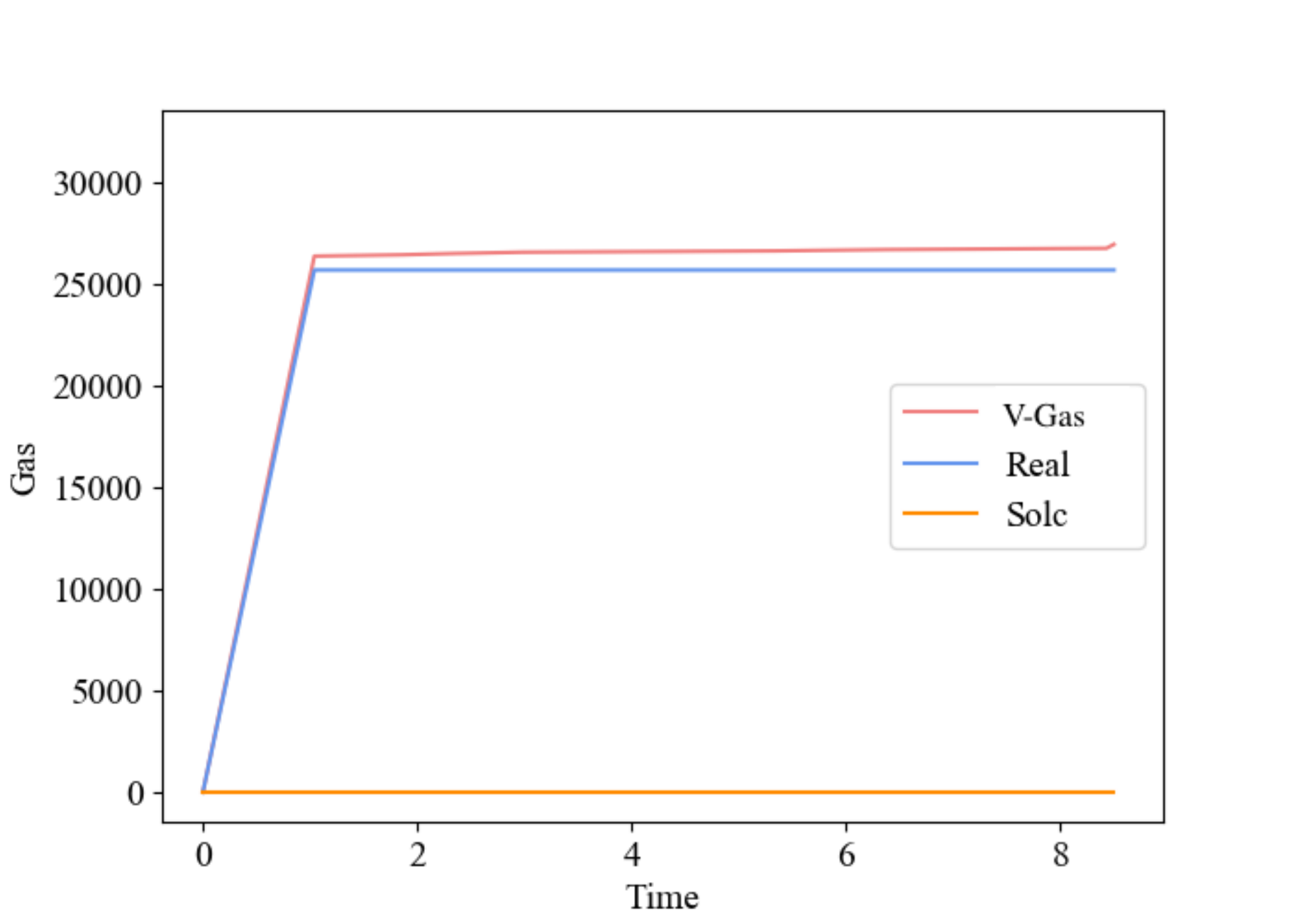}}
  \centering
  \small{(e) V-Gas on \\ ENIGMA.sol}
\end{minipage}
\hfill
\begin{minipage}{0.24\linewidth}
  \centerline{\includegraphics[width=4.0cm]{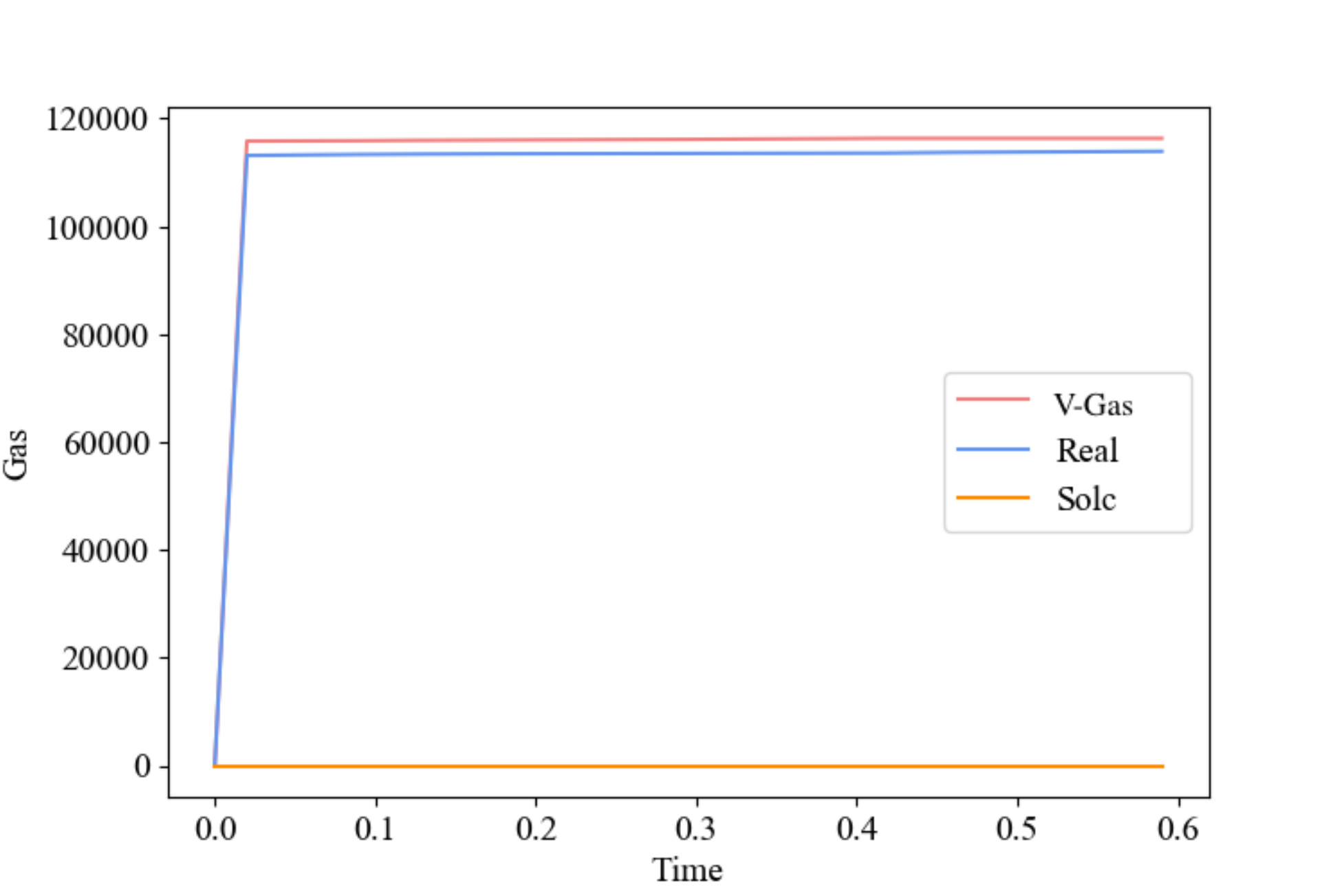}}
  \centering
  \small{(f) V-Gas on KittyItemMarket.sol}
\end{minipage}
\hfill
\begin{minipage}{0.24\linewidth}
  \centerline{\includegraphics[width=4.0cm]{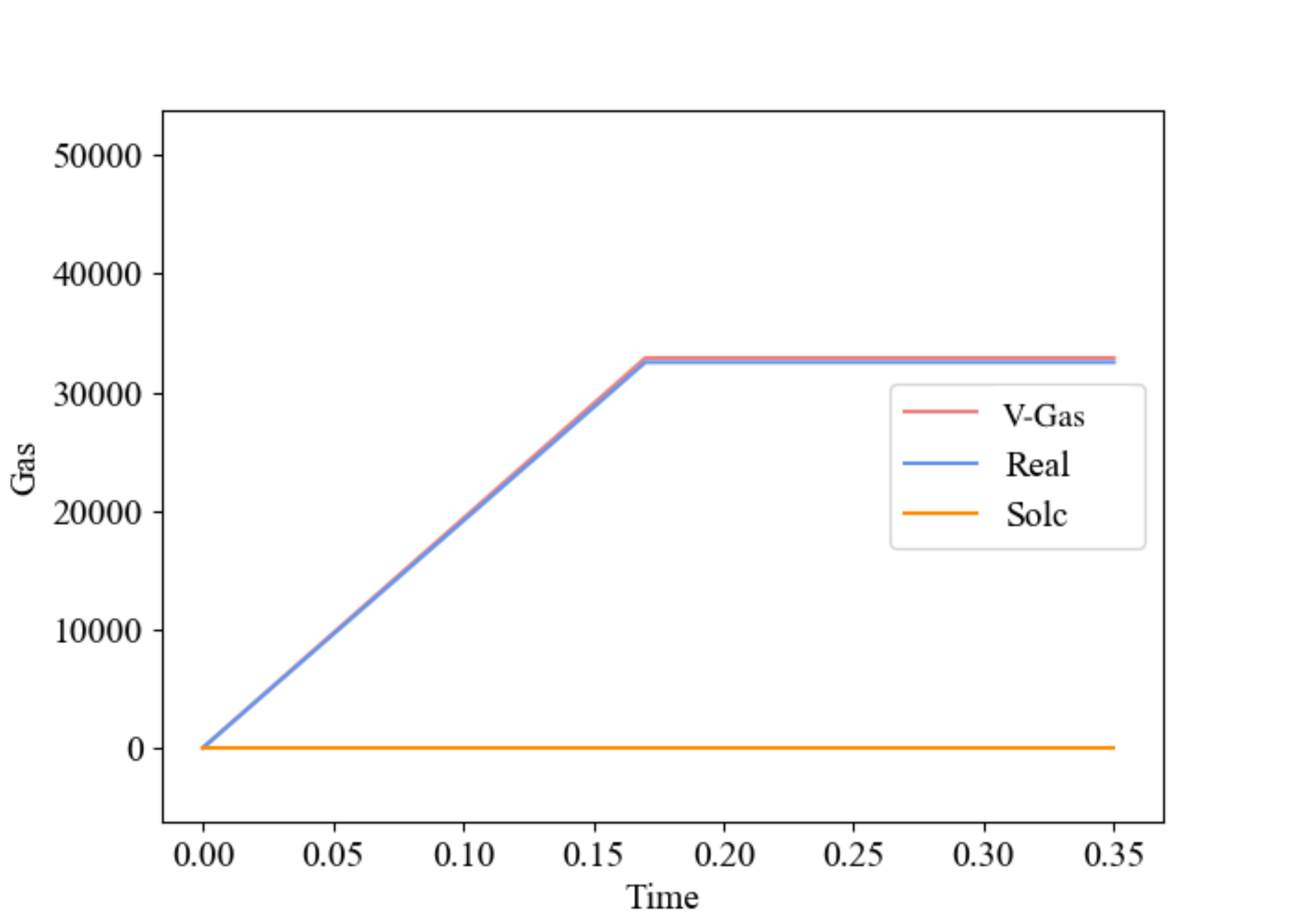}}
  \centering
  \small{(g) V-Gas on CommunityChest.sol}
\end{minipage}
\hfill
\begin{minipage}{0.24\linewidth}
  \centerline{\includegraphics[width=4.0cm]{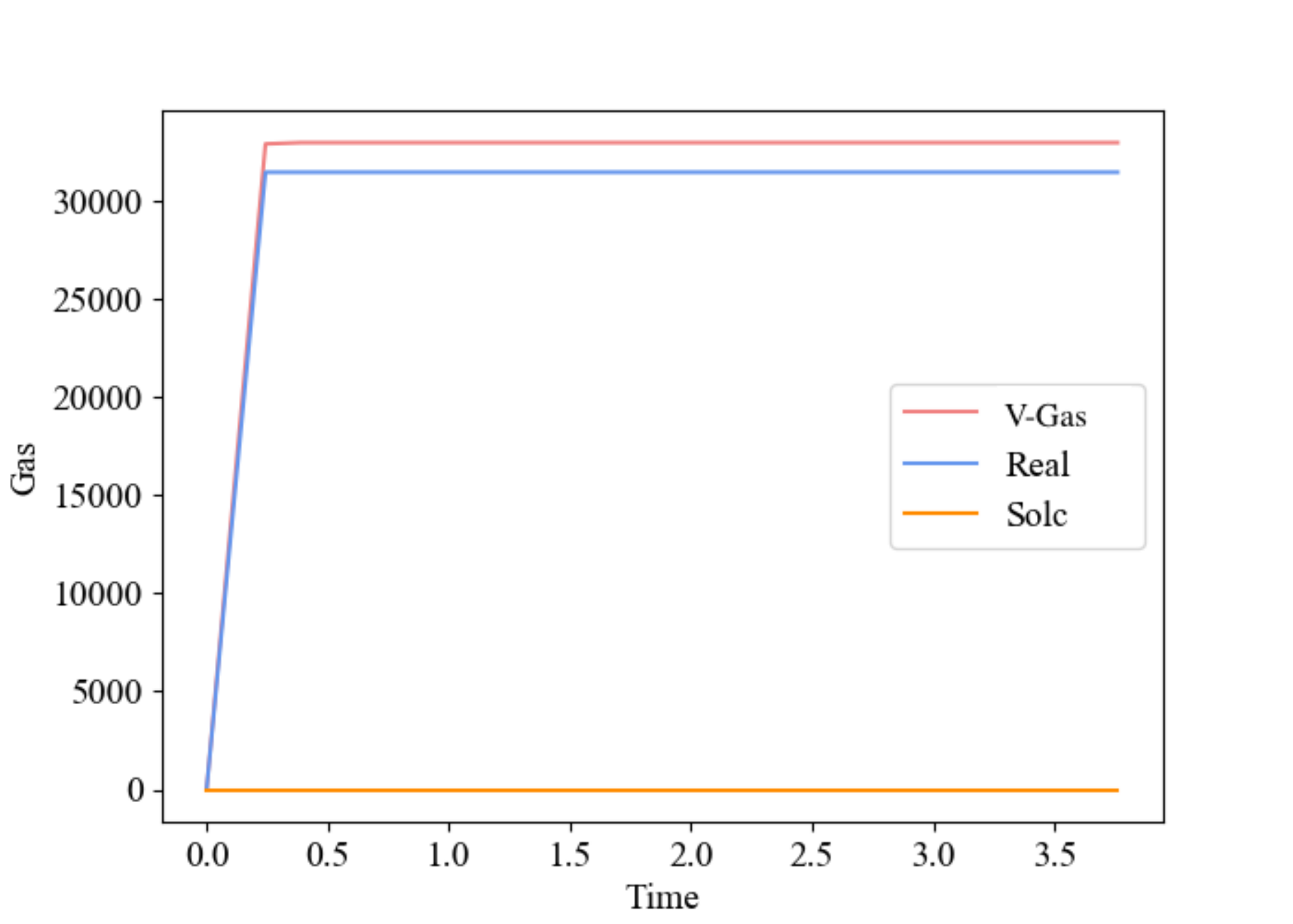}}
  \centering
  \small{(h) V-Gas on Future1Exchange.sol}
\end{minipage}
\caption{V-Gas on 8 functions. V-Gas exceeds the gas estimation results given by solc on 7 functions. V-Gas also exceeds the real maximum gas cost on 7 functions. This means with the input generated by V-Gas an out-of-gas error may occur in these transactions.}
\label{8figs}
\end{figure*}

Besides the real out-of-gas situations in these transactions, some of the transactions face potential out-of-gas errors. A potential out-of-gas means the limit of some successfully executing transactions is actually not proper in some other situations. Figure \ref{8figs} shows the relationship between the results of V-Gas on the 8 functions and fuzzing time. As the figure shows, V-Gas could approach or even exceed the real gas used in the transaction, which means some of the functions have potential our-of-gas error. With the inputs generated by V-Gas, out-of-gas errors may occur. We will use 3 functions as examples to illustrate the three types of inputs that V-Gas generated within 5 minutes as below.

\textbf{V-Gas exceeds the results of solc and the real gas cost in the transaction.} The first function we focused on is \textit{transfer} taken from the contract Dogg. This function is used to transfer some ether to some other addresses. This function needs two parameters: \textit{to} and \textit{value}. The first is the destination address and the second is the amount of tokens to be transferred. The function is shown as the following:
\lstset{language=Solidity}
\begin{lstlisting}
function transfer
(address _to, uint256 _value) public returns (bool success) 
{
    _transfer(msg.sender, _to, _value);
    return true;
}
\end{lstlisting}
'msg.sender' in function' \_transfer' represents the caller of the transaction. This a state variable used in Ethereum. V-Gas finds three different inputs for this function. At 0.32 seconds, V-Gas generates an input that could cost 25603 gas units. At 0.49 seconds, another input is generated cost 25607 units of gas. After 0.58 seconds, V-Gas finds an input to cost 55072 units if gas. This result is higher than the real gas cost in the transaction based on this function, which is 52835 units. This means there may be potential out-of-gas errors in these transactions when given the inputs generated by V-Gas. 
\\ \indent This situation occurred 28 of the 30 transactions based on eight contracts mentioned above. As for all the transactions we used in our experiment, 69.29\% of them have this situation. In this situation, if the user used solc to set the gas limit of the transaction ,there is a chance that out-of-gas error could occur. In this situation, user could just set the gas limit as a value which is a little larger than the result of V-Gas. 

\textbf{V-Gas exceeds results of solc but is lower than the real gas cost in the transaction.} The second function we focused on is 'addSERAPHIM' from the contract 'AccessoryData'. The function is listed as the following. The function is used to add a new seraphim to the system. The 'onlyCREATOR' is a modifier used to limit the scope of the executor of the function. The function first check whether the new seraphim is a new one, then it add the number of the seraphims by one.

\vspace{0.3cm}
\lstset{language=Solidity}
\begin{lstlisting}
function addSERAPHIM(address _newSeraphim) 
        onlyCREATOR public {
        if (seraphims[_newSeraphim] == false) {
            seraphims[_newSeraphim] = true;
            totalSeraphims += 1;
        }
}
\end{lstlisting}

V-Gas gives two gas consumption results for this function. At 0.135 second, V-Gas find a set of inputs that could cost 48937 units of gas. At 0.224 second, V-Gas gives another gas consumption result of 49601 units. Solc gives a result of 41921 gas units for this function. However, in the real transaction, 49816 units of gas have been consumed. V-Gas find a result better than solc but lower than the real situation.
The reason that V-Gas may not find the worst case for gas consumption may be the time for fuzzing. If we ran the tool with more time, maybe it will give us a higher result. However, this result is higher than the one of solc, which means the risks of out-of-gas error are decreased with the help of V-Gas. 
\indent \textbf{V-Gas exceeds the real gas cost but is lower than the estimation result by solc.} The last function we focused on is \textit{transfer} in BAToken.sol. This function is listed as follows. 
\lstset{language=Solidity}
\begin{lstlisting}
function transfer
(address _to, uint256 _value) returns (bool success)
{
  if (balances[msg.sender] >= _value && _value > 0) 
  {
    ...
    Transfer(msg.sender, _to, _value);
    return true;
  }
  else 
    return false;
}
\end{lstlisting}
Solc predicts that 43517 gas units would be charged while V-Gas gives a result of 25879 gas units which is lower than the estimation of solc. To find out the reason for the under estimation, we made a thorough analysis of this result. The following codes are extracted from the opcode sequences of the execution.
\begin{lstlisting}
{"pc":1420,"op":85,"gas":"0xffffaa6c","gasCost":"0x0",
"stack":["0xd356a28b","0x1f8","0x0","0x1","0x1000000000000
0000000000000000000000000000","0x0"],
"depth":0,"opName":"SSTORE"}
\end{lstlisting}
'SSTORE' is an opcode used to store a data in the memory. The opcode takes two parameters from the stack. The first element is the value that needs storing. The second element is the address that the value is stored at. As for this example, the value  '0x10000000000000000000000000000000000000000' is the address and the value '0x0' is the value stored at this address. As stated in the Ethereum yellow paper\cite{yellowpaper}, the SSTORE opcode will give a gas refund if it changes some non-zero value to a zero value in some address. The refund gas is half the gas used before, which could not exceed 15,000. In our example, the gas used before is 43,517. So the refund gas is 15,000. solc gas estimation did not consider the refunding, which leads to a higher gas estimation. 
In this situation, V-Gas gives a more accurate result than solc which is also higher than the real transaction costs. As we mentioned in Section II, fewer gas limit settings mean faster execution. Besides, excessive gas estimates are also dangerous because attackers can exploit the gas loss.
\begin{mdframed}
\textbf{Answer to RQ-b:}  The out-of-gas errors could be effectively avoid with the help of V-Gas. V-Gas is capable of generating inputs that approximate or exceed the real gas cost of the transaction in smart contract, and reduce the under estimation of solc dramatically.
\end{mdframed}

\subsection{V-Gas Engine}
V-Gas is a feedback-directed fuzzing method. However, some other feedback-directed fuzzing tools could also be easily used in this problem. We compared V-Gas with two other fuzzing tools: SlowFuzz \cite{slowfuzz} and PerfFuzz \cite{Lemieux2018PerfFuzzAG}. SlowFuzz provides a fuzzing method to detect algorithmic complexity vulnerabilities for C/C++ programs. SlowFuzz uses resource-usage-guided evolutionary search techniques to automatically find inputs that maximize computational resource utilization for a given application. Similar to SlowFuzz, PerfFuzz generates inputs via feedback-directed mutational fuzzing. PerfFuzz uses multi-dimensional feedback and independently maximizes execution counts for all program locations. 

\subsubsection{Implementations for Gas Used Problem.} In order to apply the fuzzing tools mentioned above, based on their open-source version, we customize their feedback mechanisms for fuzzing smart contracts. SlowFuzz retains the seeds that could maximize the total path of program execution. While PerfFuzz retains seeds that could lead to more execution of an edge in the program CFG. In our customization, we use the same CFG as V-Gas. For SlowFuzz, we calculated the execution times for each edge in the CFG during an execution. And the total length of path is represented as the sum of execution times for all edges. As for PerfFuzz, we record the maximum execution times for each edge in the CFG. If a seed could execute some edges more than the times recorded as the maximum value, it will be reserved.

\subsubsection{Comparison among Different Fuzzing engines.} We compared V-Gas with the fuzzing tools as well as a random fuzzing engine. The random engine save all the seeds randomly without any feedback information. The comparison is committed on the 736 transactions we took. The result of the comparison is shown in Table\ref{max_gas}.

\begin{table}[H]
\caption{The number of transactions that each fuzzer can maximize the gas estimation compared with other engines}
\label{max_gas}
\begin{center}
\begin{tabular}{c|cccc}
\hline
        & PerFuzz & RnadomFuzz & SlowFuzz& V-Gas  \\ \hline

total      & 628  & 644  & 623 &  \textbf{722}    \\ 
percentage & 85.1\%  & 87.3\% & 84.4\% & \textbf{97.8\%}    \\ \hline
\end{tabular}
\end{center} 
\end{table}

\begin{table*}[!htbp]
\caption{Performance of each fuzzer on 8 smart contracts}
\label{gasVsothers}
\begin{center}
\begin{tabular}{c|ccc|ccc|ccc|ccc}
\hline
\multirow{2}{*}{} &
\multicolumn{3}{c|}{PerFuzz} &
\multicolumn{3}{c|}{RandomFuzz} &
\multicolumn{3}{c|}{SlowFuzz} &
\multicolumn{3}{c}{V-Gas} \\
\cline{2-13}
  & Result(unit) & Time(s) & GasRate
  & Result(unit) & Time(s) & GasRate
  & Result(unit) & Time(s) & GasRate
  & Result(unit) & Time(s) & GasRate \\
\hline
BAToken & 25879 & 0.712 & 36347
        & 25879 & 3.012 & 8592
        & 25879 & 2.166 & 11948
        & 25879 & 0.542 & \textbf{47747}\\
DSTokenBase & 53179 & 2.412 & 22048
            & 53179 & 1.904 & 27930
            & 53179 & 0.579 & \textbf{91846}
            & 53179 & 1.949 & 27285\\
Dogg & 55072 & 4.457 & 12356
     & 25667 & 2.287 & 11223
     & 55008 & 2.549 & 21580
     & 55072 & 0.583 & \textbf{94463}\\
AccessoryData & 49601 & 0.704 & 70456
              & 49601 & 2.315 & 21426
              & 49601 & 1.467 & 33811
              & 49601 & 0.224 & \textbf{221433}\\
ENIGMA & 26949 & 52.246 & 516
       & 26693 & 60.543 & 441
       & 26565 & 3.751 & \textbf{7082}
       & 26949 & 8.496 & 3172\\
KittyItemMarket & 116196 & 2.522 & 46073
                & 116196 & 0.546 & 212813
                & 116324 & 1.681 & 69199
                & 116324 & 0.417 & \textbf{278954}\\
CommunityChest & 32884 & 0.426 & 77192
               & 32884 & 158.368 & 208
               & 32820 & 0.651 & 50415
               & 32884 & 0.173 & \textbf{190081}\\
Future1Exchange & 32991 & 1.234 & 26735
                & 32991 & 1.601 & 20606
                & 32991 & 0.767 & 43013
                & 32991 & 0.389 & \textbf{84810}\\
\hline
\end{tabular}
\end{center} 
\end{table*}

\begin{figure}[hbp]
\centering
\includegraphics[height=6.0cm, width=8.0cm]{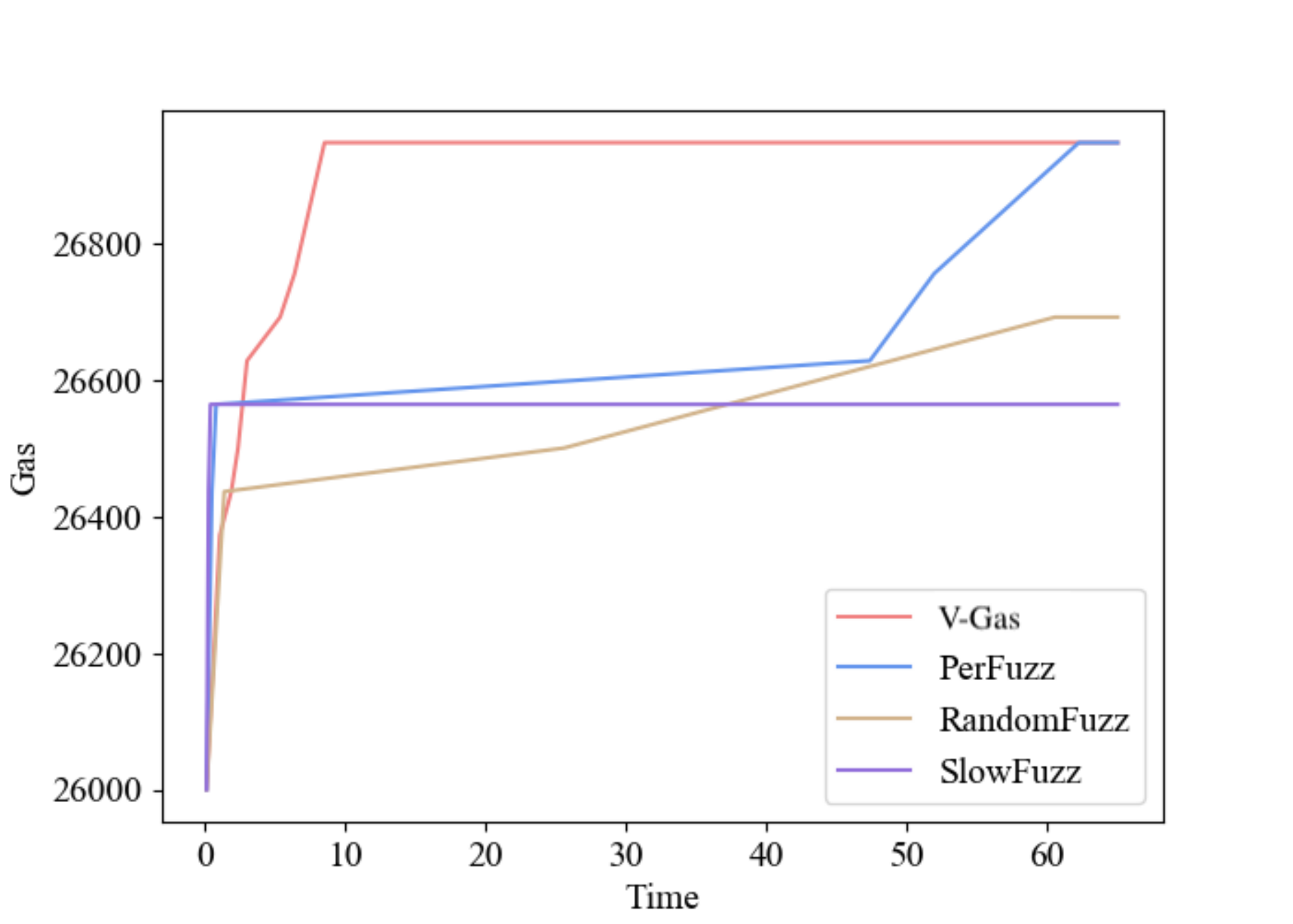}
\caption{Trend of results given by 4 fuzzing engines on function 'Activate' from contract 'ENIGMA'}
\label{ENIGMALine}
\end{figure}

V-Gas performs the best in 97.8\% of the transactions compared with other fuzzing engines, which is the best among all of the engines. RandomFuzz performs better than the other two fuzzing engines with a ratio of 87.3\%. The reason for this result may be that the directed strategies used in Perfuzz and SlowFuzz are not suitable for this problem. Maximizing the execution times for each edge or the total length of the execution path may lead to a highest gas cost brought by the opcodes. However, the gas consumption of a transaction also contains the gas cost brought by data storing. For this reason, the fuzzing engines in PerFuzz and SlowFuzz are not ideal in this problem. In order to observe the difference among the four engines more specifically, we count the fuzz results of the four engines on the eight contracts we mentioned above. The results are shown in Table\ref{gasVsothers}. In the table, we introduce a new concept called 'GasRate'. Its definition is shown in the following formula:
\[
\begin{split}
\ GasRate = Result / Time \
\end{split}
\]
This variable is used to measure the effectiveness of the fuzz engine. If this tool can find higher gas consumption in less time, which means a higher GasRate, we think the effectiveness of this engine is better than others. As the table shows, V-Gas has a highest GasRate on 6 contracts. V-Gas finds the highest gas cost among four engines on all of the 8 contracts. V-Gas can find the best results in a relatively short period of time. We now use the contract 'ENIGMA' to show the fuzzing process of these fuzzing engines. We chose the contract 'ENIGMA' because the 4 engines gives many gas consumption results of function 'Activate' in this contract which could clearly show the trend of the results of the 4 engines in the process of fuzzing. The results of 4 fuzzing engines are shown in Fig\ref{ENIGMALine}.

As we could see from the figure, V-Gas could reach a high result in a short time which is 8.5 seconds to be precisely. PerFuzz gives the same result as V-Gas at 52.24 seconds. SlowFuzz and RandomFuzz performs worse than the other two engines even with more than 5 minutes. 

\begin{mdframed}
\textbf{Answer to RQ-c:} V-Gas performs better than some well used feedback-directed fuzzing engines in this problem, where the gas weighted control flow graph and the gas directed seed selection and mutation strategies contribute to the improvements.
\end{mdframed}

\subsection{Practical Applications}
V-Gas aims at generating a set of inputs that could trigger a high gas cost. It can be useful for different practical purposes, and we evaluate V-Gas for two potential applications. 

\subsubsection{Gas Estimation and Limit Setting}
Choosing a proper gas limit for each transaction is a tough problem. Many works introduced in Section 7 focus on solve this problem in different ways. V-Gas could also be used to estimate the gas limit of a transaction. To our best knowledge, except for solc, we are the second publicity available tool 
\footnote{http://github-xxxx-xxx, hidden for double blind review and would be open after review process}. The popular solc often generates `infinite' estimation or under estimation. V-Gas could give a more accurate result as presented in Table \ref{Ability}.

\subsubsection{Gas Related Vulnerabilities Detection}
Another application of V-Gas is to detect gas related vulnerabilities. A transaction with a gas cost that exceeds the gas limit will be reverted. Malicious attackers may exposure a set of inputs to trigger a loop to consume plenty of gas. As a result, the service provided by the contract is stopped. Such vulnerabilities may cause a loss of money. The following code 
\lstset{language=Solidity}
\begin{lstlisting}
function transfer
(address[] _address, uint256[] _values) onlyOwner public {
 require(_address.length == _values.length);
 for (uint i = 0; i < _address.length; i += 1) {
        token.transfer(_address[i],_values[i]); }
 AirDroppedTokens(_address.length);
}
\end{lstlisting}
is fetched from a real-world smart contract. This function is used to transfer some ether to some accounts. Attackers may create a bunch of address to cause a denial of service attack to this contract. The transfer function in the loop may be executed for many times and cost plenty of gas. As a result, the gas used exceed the gas limit of a block and the transaction would revert.

V-Gas could generate a set of inputs which identifies this problem. This previously unknown issue has been assigned with a CVE ID. We have already detected 25 confirmed out-of-gas vulnerabilities with the help of V-Gas, 5 of which have been assigned unique CVE identifiers in the US National Vulnerability Database and other 20 are appended for approval.

\begin{mdframed}
\textbf{Answer to RQ-d:} V-Gas has practical values and is really helpful in gas estimation tasks as well as in looking for gas-related vulnerabilities.
\end{mdframed}




\section{Threats to Validity}



\textbf{Mutation and selection strategies limitations.} The mutation strategies of V-Gas are hybrid for the current version. They are designed based on the mutation strategies in AFL. V-Gas uses some global functions to represent the environment variables. For mutations, these variables are given different values randomly on the basis of their types. This mutation strategy performed correctly for V-Gas. However, more efficient mutation strategies for environment variables could be adapted to V-Gas to build a more complex environment for the contract to run. A more complex environment means more accounts and more state information, it may lead to some different behaviours of smart contracts. Second, even we have implemented MCMC in the seed selection, we will still face the local optimization problem, which needs to be solved with more efficient algorithms or improve the reservation probability for none-interesting seed .

\textbf{Smart contracts support limitations.} The current prototype of V-Gas could not deal with the smart contract functions that call another contract  deployed on some address in Ethereum. The gas cost of these functions determined by the called contract, which could not be calculated by V-Gas for now. This problem may be solved by fetching the called contract and making a relationship between these two contracts. We will try to make V-Gas adapted to these kinds of contracts in our future work.

\textbf{Data collection limitations.} One of the subjective threats is the data collection of the experiment. The experimental process contains a large amount of data, and there are often some errors in the statistics and analysis phases. In order to solve this problem, we use fully automated script files to statistics and analyze the data. To a certain extent, script files could avoid the impact of the data collection process on the experimental results. Another possible subjective threats of this experiment is the time of fuzzing process. The result of fuzzing technique depends on the duration time sometimes. More fuzzing time may give a larger result of gas costs. 
\section{Related Work}

\textbf{Gas cost estimation.}  Solc \cite{solc} provides an api to estimate the gas cost of the functions of contracts as showed in Section 5. However, the estimation way of solc has a lot of constraints which lead to many `infinite' estimations and under estimations. GASTAP \cite{Albert2018GASTAPAG} infers sound gas upper bounds for all public functions in a smart contracts with complex transformation and analysis processes on the code. Matteo Marescotti \cite{marescotti2018computing} have presented a solution to the problem of estimating the gas consumption for Ethereum smart contracts based on techniques inspired by bounded model-checking techniques. These works contribute lots of premiere ideas for further research, but none are publicity available except for Solc.

\textbf{Security of Ethereum and Smart contracts.} Security of Ethereum and smart contracts is essential to the development of blockchain technique. Some works make a survey on the attacks of Ethereum and smart contracts \cite{atzei2016survey}. Plenty of research focus on Ethereum and smart contract security. For example, echidna \cite{echidna} provides a fuzzing framework for developers to check if there is any problem in their contracts. It takes a list of invariants (properties that should always remain true) as input. For each invariant, it generates random sequences of calls to the contract and checks if the invariant holds. 
ContractFuzzer \cite{contractfuzzer} defines some oracles to detect vulnerabilities in smart contracts such as dangerous delegate bugs. Vandal \cite{brent2018vandal}  presents a security analysis framework for Ethereum smart contracts. 
Another recent work\cite{grishchenko2018semantic} gives the first complete small-step semantics of EVM bytecode. This formalization relies on a combination of hyper and safety properties. MadMax \cite{grech2018madmax} classifies and identifies gas-focused vulnerabilities, and present a static program analysis technique to automatically detect gas-focused vulnerabilities with very high confidence.  EVMFuzz \cite{fu2019evmfuzz} uses differential testing to fuzz different versions of EVM and try to find out the vulnerabilities in EVM. Our another work EVM* provides a framework to make a reinforcement on EVM \cite{Ma2019EVMFO} to test whether there is a dangerous operation during transaction execution and stop the dangerous transaction in time.

\textbf{Performance problems detection.} A lot of work has been done to detect the waste of gas and the optimazation pattern for gas consumption. For example, a recent work \cite{chen2018towards} identifies 24 anti-patterns from the execution traces of real smart contracts. Besides, they design and develop GasReducer, the first tool to automatically detect all these anti-patterns from the bytecode of smart contracts and replace them with efficient code through bytecode-to-bytecode optimization. Their another work \cite{chen2017adaptive} proposes an emulation-based framework to automatically measure the resource consumptions of EVM operations.

\textbf{Fuzzing tools.} Traditional fuzzing techniques aims at finding memory safety problems. Fuzzing testing could be divided as mutation based fuzzing and generation based fuzzing. Mutation based fuzzing such as AFL\cite{AFL}, mutates seeds with mutation strategies for each fuzzing process. Generation based fuzzing generate inputs based on the domain knowledge of the program. Some works focus on the improvement of AFL to increase the fuzzing performance. Enfuzz\cite{chen2018enfuzz} proposes a method to combine several fuzzers to work together and share the seeds with each other. Pafl\cite{liang2018pafl}  utilizes efficient guiding information synchronization and task division to extend those existing fuzzing optimizations of single mode to industrial parallel mode. Safl\cite{wang2018safl} focuses on the combination of symbolic execution and fuzzing testing tools to increase the coverage. 
Some other works focus on the techniques to find the weakness of fuzzing techniques, such as Anti-fuzz\cite{Edholm2016EscapingTF}. Their results show that it is relatively easy to implement and apply anti-fuzzing techniques that are able to completely mask crashes and, by extension, vulnerabilities from fuzzers. However this is hard to implement in smart contracts deployed on Ethereum, because the contracts on Ethereum could not be changed since it is deployed.

\textbf{Main difference.} For the gas estimation tools, V-Gas focuses on generating inputs to trigger a high gas consumption. Besides, in section V, we demonstrated V-Gas could perform better than the related avaiable tools.  For the traditional fuzzing tools aimed at smart contracts, V-Gas uses different mutation strategies and detect different vulnerabilities. Echidna and ContractFuzzer randomly mutates the inputs for each generation, while V-Gas combines the strategies of AFL and the randomly methods for environment variables. Besides, V-Gas aims at generating inputs trigger a high gas cost but not detecting the memory safety bugs. As for the performance bug detection tools, V-Gas could automatically find the bottleneck of gas cost for a smart contract, either for gas limit reference or vulnerability detection. The previous works aim at C/C++ programs or artificially define some optimization or known patterns for smart contracts.

\section{Conclusion}
In this paper, we designed V-Gas, a fuzzing approach to generate inputs that could lead to a high gas consumption, which can be used to reduce the under estimation ratio of existing tools and the detect previously unknown out-of-gas vulnerabilities. First, V-Gas will generate a W-CFG for the function. Second, V-Gas uses feedback-directed selection and mutation strategies. Finally, V-Gas provides an environment for contract transaction.
The result demonstrates that V-Gas could successfully generate some inputs to trigger high gas cost in a short time.
In 86.14\% of the transactions, the gas cost given by V-Gas is higher than the gas cost of real-world transactions or values estimated by solc, which could successfully decrease the risk of out-of-gas error, from 44.02\% to 13.86\% of all transactions. 
For the practical application of V-Gas, we evaluate V-Gas on more real-world contracts and detect 25 previously unknown gas-related vulnerabilities, and 5 of which are assigned with CVE Ids.

Our future work will focus on the optimization of V-Gas. 
We will design some gas-related mutators, accelerate the fuzzing procedure with some static analysis such as the result of solc, and try to detect more gas-related vulnerabilities in smart contracts with V-Gas. Besides, we would like to find some gas hot spots with the help of V-Gas and try to design some patterns to optimize the gas cost of these spots.

\bibliographystyle{IEEEtran}
\bibliography{IEEEabrv,GasFuzz}

\end{document}